\documentclass[preprint,notoc]{JHEP3}
\usepackage{epsfig}

\def\etmiss{E_T^{\rm miss}}
\def\eslt{E_T^{\rm miss}}

\def\to{\rightarrow}

\def\bi{\begin{itemize}}
 \def\ei{\end{itemize}}
\def\te{\tilde e}
\def\shat{\hat{s}}
\def\c1p{C1^\prime}

\def\tst{\tilde t}
\def\ttau{\tilde \tau}
\def\tmu{\tilde \mu}
\def\tg{\tilde g}
\def\tnu{\tilde\nu}

\def\tq{\tilde q}

\def\tw{\widetilde W}
\def\tz{\widetilde Z}
\def\alt{\stackrel{<}{\sim}}
\def\agt{\stackrel{>}{\sim}}
\def\be{\begin{equation}}  
\def\ee{\end{equation}}  
\def\bea{\begin{eqnarray}}  
\def\eea{\end{eqnarray}}  

\def\sps1ap{SPS1a$^\prime$}
\title{Supersymmetry discovery potential of the LHC at\\
$\sqrt{s}=$10 and 14 TeV without and with missing ${\bf{E_T}}$ }
\author{Howard Baer$^{a}$, Vernon Barger$^b$, Andre Lessa$^a$ 
and Xerxes Tata$^{c}$\\
$^a$Dept.\ of Physics and Astronomy, University of Oklahoma, Norman, OK 73019, USA\\
$^b$Dep't of Physics, University of Wisconsin, Madison, WI 53706, USA\\
$^c$Dept. of Physics and Astronomy, University of Hawaii, Honolulu, HI 96822, US\\
E-mail: \email{baer@nhn.ou.edu}, \email{barger@physics.wisc.edu},
\email{lessa@nhn.ou.edu}, \email{tata@phys.hawaii.edu}}

\preprint{\vbox{UH-511-1142-09}}

\abstract{ We examine the supersymmetry (SUSY) reach of the CERN LHC
operating at $\sqrt{s}=10$ and 14 TeV within the framework of the
minimal supergravity model (mSUGRA). We improve upon previous reach
projections by incorporating updated background  calculations
including a variety of $2\to n$ Standard Model (SM) processes. We
show that SUSY discovery is possible even before the detectors are
understood well enough to utilize either $\eslt$ or electrons in the signal.
We evaluate the early SUSY reach
of the LHC at $\sqrt{s}=10$ TeV by examining multi-muon plus $\ge
4$ jets, and also lepton-free, acollinear dijet events with {\it no} missing $E_T$ cuts,  
and show that the greatest reach in terms of $m_{1/2}$ occurs in the
dijet channel, where it may be possible to probe $m_{\tq}\sim m_{\tg}
\alt 1$~TeV with just 1~fb$^{-1}$ of integrated luminosity.  The reach
in multi-muons is slightly smaller in $m_{1/2}$, but extends to higher
values of $m_0$.  We find that an observable multi-muon signal will
first appear in
the opposite-sign dimuon channel, but as the integrated luminosity
increases the relatively background-free but rate-limited same-sign
dimuon, and ultimately the trimuon channel yield the highest reach.  
The optimized reach in these channels extends to $m_{\tg} \alt 600$
(800)~GeV for an integrated luminosity of 100~pb$^{-1}$ (1~fb$^{-1}$). 
We
show characteristic distributions in these channels that serve to
distinguish the signal from the SM background, and also help to
corroborate its SUSY origin. We then evaluate the LHC reach in various
no-lepton and multi-lepton plus jets channels {\it including} missing
$E_T$ cuts for $\sqrt{s}=10$ and 14 TeV, and plot the reach for
integrated luminosities ranging up to 3000 fb$^{-1}$ at the SLHC. For
$\sqrt{s}=10$ TeV, the LHC reach extends to $m_{\tg}=1.9,\ 2.3,\ 2.8$
and $2.9$ TeV for $m_{\tq}\sim m_{\tg}$ and integrated luminosities of
10, 100, 1000 and 3000 fb$^{-1}$, respectively. For $\sqrt{s}=14$ TeV,
the LHC reach for the same integrated luminosities is to $m_{\tg}=2.4,\
3.1,\ 3.7$ and $4.0$ TeV, respectively. The reach estimates for ab$^{-1}$
luminosities may be over-optimistic due to low statistics of background
with very hard cuts.
 }  

\begin{document}

\section{Introduction}
\label{sec:intro}

The CERN Large Hadron Collider (LHC) is expected to begin collecting
data from $pp$ collisions at $\sqrt{s}\sim 10$~TeV, 
with a goal of accumulating $0.1-0.2$ fb$^{-1}$ of usable data in the
first run.  During the very early stages of LHC running (first $\sim
0.1$ fb$^{-1}$), detector commissioning will be in progress, and issues
such as detector alignment and calibration will be addressed, as the
experimental groups use familiar Standard Model (SM) processes such as
$W+jets$, $Z+jets$ and $t\bar{t}$ production to guide the way. Running
at 10~TeV is likely to continue for a year or more, after which it is
expected that the center of mass energy will be increased, very likely
in several stages, to its design value of 14~TeV.

While discovery of the Higgs boson (or bosons) or, more generally, the
mechanism of electroweak symmetry breaking remains a primary goal of LHC
experiments, an integrated luminosity of $\sim 10$~fb$^{-1}$ will be
required to claim Higgs discovery (if indeed $m_{\rm Higgs}\sim 115-130$
GeV, as indicated by global analyses of electroweak data sets)
\cite{mhiggs}.  An equally important objective for LHC is to discover,
or exclude, weak scale supersymmetric (SUSY) matter. Since production
cross sections for strongly interacting sparticles can range up to
$\cal{O}$$(10^5)$ fb if $m_{\tq}\sim m_{\tg} \sim 400$ GeV at $\sqrt{s}=10$~TeV,
the hunt for supersymmetric particles beyond
the reach of LEP2 and Tevatron searches could be very interesting even
in the earliest stages of LHC running.

The discovery capability of LHC for SUSY particles is often illustrated
with a reach plot in the parameter space of some assumed SUSY
model \cite{wss}.  At each point in SUSY model parameter space, many
simulated collider events are generated, and compared against SM
backgrounds with the same experimental signature\cite{bg}. 
Judicious cuts are
then  implemented to select out the new physics signals over SM
backgrounds, and the signal is deemed observable if it satisfies
pre-selected criteria for observability.
%
The LHC reach has most frequently been analyzed \cite{lhcreach} within
the paradigm minimal supergravity (mSUGRA) model \cite{msugra}, but
other SUSY models have also been studied.\footnote{The LHC reach in
anomaly-mediated SUSY breaking (AMSB) models is given in
Ref.~\cite{amsb}, in the mixed-modulus-anomaly mediation case in
Ref.~\cite{mmamsb}, while reach for various model lines in gauge
mediated SUSY breaking (GMSB) is presented in Ref. \cite{gmsb},
and for gaugino-mediated SUSY breaking in Ref. \cite{gaugino}. The LHC
reach in mSUGRA with $R$-parity violation is presented in
Ref. \cite{rviol}.}

In many models, production of strongly interacting SUSY particles is
expected to yield the dominant signal channel, at least for
$m_{\tg}\simeq m_{\tq}\alt 1.7$~TeV. Heavy squarks and gluinos then decay
via complex cascades\cite{cascade} which, 
if $R$-parity conservation is assumed, ends
in the stable (or quasi-stable) lightest SUSY particle (LSP), 
often assumed to be the lightest neutralino $\tz_1$.  
The $\tz_1$ escapes experimental detection, so that the generic SUSY
signal is expected to be the production of multiple high $E_T$ jets,
multiple high $p_T$ isolated leptons ($e$ or $\mu$, produced via the
decays of chargino and neutralino secondaries) and possibly also isolated
photons, together with missing transverse
energy $\eslt$.  The multiplicity of isolated leptons provides a
convenient way to classify various SUSY signals \cite{btw}, and for the mSUGRA
model, reach contours have been shown for signals in the following channels:
\begin{itemize}
\item $jets+\eslt$ (no isolated leptons),
\item $1\ell +jets+\eslt$,
\item two opposite-sign isolated leptons (OS)$+jets+\eslt$,
\item two same-sign isolated leptons (SS)$+jets+\eslt$,
\item $3\ell +jets+\eslt$,
\item $4\ell +jets+\eslt$,
\item a real $Z\to\ell^+\ell^- +jets +\eslt$,
\item a hard, isolated $\gamma +jets +\eslt$.
\end{itemize} 
These explorations have typically been performed for the design
LHC center-of-mass energy of $\sqrt{s}=14$ TeV, and 
integrated luminosities of 10 or 100 fb$^{-1}$, anticipated after 
a year to a few years of running at the design luminosity.

Recently, some attention has been given to the ability of LHC to detect
supersymmetric matter in the very earliest stages of running when 
a reliable measurement of $\eslt$, which requires a lead-time for
detector alignment, calibration and understanding of the performance of
essentially all detector components, will not be available \cite{green}. 
In Ref. \cite{bps} it was instead suggested that in lieu of $\eslt$,
high isolated lepton multiplicity could be used as a strong cut to
reject SM backgrounds at relatively low cost to the expected SUSY signal: thus,
requiring events with $\ge 4$ jets plus OS, or SS, or three isolated
leptons could allow for probes of $m_{\tg}\sim 500-600$ GeV with just
0.1-0.2 fb$^{-1}$ of integrated luminosity.  In a follow-up \cite{bls}
paper, the authors refined and restricted their multi-lepton analysis to just
multi-muons (because reliable electron identification may be difficult in
the early stages of LHC running). The LHC reach was evaluated for
$\sqrt{s}=10$ TeV, and was found to be $m_{\tg}\sim 550$ GeV for 0.2
fb$^{-1}$ in the SS dimuon plus $\ge 4$ jets channel.\footnote{Indeed,
muons are already being seen by ATLAS and CMS in cosmic ray events, and
further, muons can be readily identified at lower $p_T$ values
than electrons, thus partially compensating for the loss of electron
channel due to the increased muon signal efficiency.}  Alternatively, a
search for acollinear dijet events  was
suggested by Randall and Tucker-Smith (RT-S)\cite{rts} as a SUSY search
strategy that did not explicitly require $\eslt$. By cutting hard
on several variables, a signal detectable over SM background was found, 
especially 
over portions of mSUGRA parameter space where squark pair production is 
significant and where the squarks decay directly into $q\tz_1$.

In this paper, where we re-assess the LHC SUSY reach within
the mSUGRA model, we have several goals: 
\begin{enumerate}
\item We perform much more
detailed SM background calculations than many previous works, including
many $2\to n$ subprocesses.  In processes such as $W$, $Z$ and
$t\bar{t}$ production, we include exact parton emission matrix elements
for the first several quark or gluon radiations (see details below).
These calculations should model the multiple high $E_T$ jet production
in association with standard processes to much better accuracy than the
parton shower method. In addition, we include numerous other
subprocesses such as $Zt\bar{t}$ and $t\bar{t}b\bar{b}$ production,
which have frequently been neglected. 

\item  We evaluate the LHC early
discovery reach without $\eslt$ cuts in two additional multi-muon
channels -- OS dimuons and trimuons-- that have not yet been presented.
We also show various distributions associated with these quantities that
should lead to increased confidence that any observed excess
arises from a real signal.  We also evaluate the mSUGRA
reach in the RT-S dijet channel, and compare with the multi-muon reach.

\item We present reach plots for the initial energy option of
$\sqrt{s}=10$~TeV, and compare with similar reach plots for
$\sqrt{s}=14$~TeV. We are motivated to do so because just how and when
the center-of-mass energy of the LHC will be increased to its design
value is presently unclear: if running full current through the
superconducting magnets is deemed dangerous, or if it is deemed
impractical to re-train the magnets that have ``lost training'' so that
these cannot attain the full field, then it may be the case that LHC
runs below the design energy for the first several years.

\item We show the LHC SUSY reach for a wide range of integrated
luminosities, ranging from 0.05 fb$^{-1}$ (with cuts pertinent to early
reach) up to 3000 fb$^{-1}$ with cuts optimized for the extraction of
the SUSY signal.  These high luminosity values would only be accessible
at the SLHC, which is intended to upgrade the LHC luminosity to
$\mathcal{L}=10^{35}$ cm$^{-2}$s$^{-1}$. We stress that several
experimental challenges at such high luminosities would have to be
overcome and our background MC might not be considered realistic for
such luminosities.\footnote{For instance, multiple scattering effects
will have to be accounted for.}  Our results for 1000 fb$^{-1}$ and 3000
fb$^{-1}$ are intended to provide an outer limit of the SLHC reach for
its first few years of running. 
With this in mind, we show
a summary of the LHC/SLHC reach for integrated luminosity
values of 10, 100, 1000 and 3000 fb$^{-1}$ in
Table~\ref{table:reach}.
\end{enumerate}

\begin{table}
\centering
\begin{tabular}{|c|c|c|c|c|}
	\hline
& 10~fb$^{-1}$ &  $100$ fb$^{-1}$  & $1000$ fb$^{-1}$ & $3000$ fb$^{-1}$   \\
	\hline
$\sqrt{s}=$ 10 TeV  & 1.9~TeV  & 2.3 TeV & 2.8 TeV & 2.9 TeV \  (oFIT)\\
$\sqrt{s}=$ 14 TeV  & 2.4~TeV & 3.1 TeV & 3.7 TeV & 4.0 TeV \ (oFIT)\\
	\hline
\end{tabular}
\caption[]{Reach for the gluino mass for integrated luminosity values of
10, 100, 1000 and 3000 fb$^{-1}$ at $\sqrt{s}=$ 10 TeV and 14 TeV, assuming
$m_{\tq}
\sim m_{\tg}$. The numbers for 1000~fb$^{-1}$ and 3000~fb$^{-1}$ should
really be regarded as upper limits on the SLHC reach. For
more details see Sec. \ref{sec:reach}.}
\label{table:reach}
\end{table}

The remainder of this paper is organized as follows.  In
Sec. \ref{sec:bg}, we present details of our improved SM background
calculations, along with plots for the total background levels at
$\sqrt{s}=10$ and 14 TeV from various SM processes.  In
Sec. \ref{sec:early}, we present {\it early} \ SUSY discovery reach
plots in the various multimuon channels, but with no $\eslt$ cut, for
several integrated luminosity values and $\sqrt{s}=10$ TeV. We show
several distributions that would serve to both distinguish the signal
from SM backgrounds as well as to make a case for its SUSY origin.  We
also evaluate the LHC SUSY reach in the RT-S dijet channel.  In
Sec. \ref{sec:reach}, we show updated LHC reach plots for standard
mSUGRA signal channels including $\eslt$ cuts and our improved
backgrounds, for $\sqrt{s}=10$ and $14$ TeV, and a wide range 
of integrated luminosities. Because $b$-jet tagging, which can
potentially increase the gluino reach by up to 20\% in the mixed
bino-higgsino LSP case that occurs in the so-called hyperbolic
branch/focus point (HB/FP) region of the mSUGRA model \cite{btag}, will
be inefficient at the early stage and is presently uncertain in the
ultra-high luminosity environment of the SLHC, we do not include it in
the present analysis.  Our ultimate plots include scans over a vast grid
of possible cut values, so signal/background is optimized in various
regions of model parameter space.  We end with a summary of our results
in Sec.~\ref{sec:conclude}.

\section{Standard model background calculations}
\label{sec:bg}

In order to understand how SUSY searches are affected by changes in the
beam energy in the distinct channels, a careful assessment of the SM
backgrounds is necessary. In particular, relaxing the $\eslt$ cut may
increase the contributions from different background processes that are
usually neglected in the literature. We used AlpGen\cite{alpgen} and
MadGraph\cite{madgraph} to compute the following $2 \to n$ processes:
$jj$, $b\bar{b}$, $W^{\pm}j$, $Z^{(*)}j, \gamma^{(*)}j$,
$b\bar{b}b\bar{b}$, $t\bar{t}$, $VV$, $b\bar{b}Z$, $b\bar{b}W^{\pm}$,
$b\bar{b}t\bar{t}$, $t\bar{t}Z$, $t\bar{t}W^{\pm}$, $VVV$, $t\bar{t}VV$,
$t\bar{t}t\bar{t}$ and $VVVV$, where $j$ stands for light partons ($u$,
$d$, $s$, $c$ and $g$) and $V=W^{\pm},Z$. The leading order (LO) total
cross-sections for these processes at $10$ TeV and $14$ TeV are shown in
Fig. \ref{fig:xsecs10} and \ref{fig:xsecs14}.  
\FIGURE[t]{
\includegraphics[width=11cm,clip]{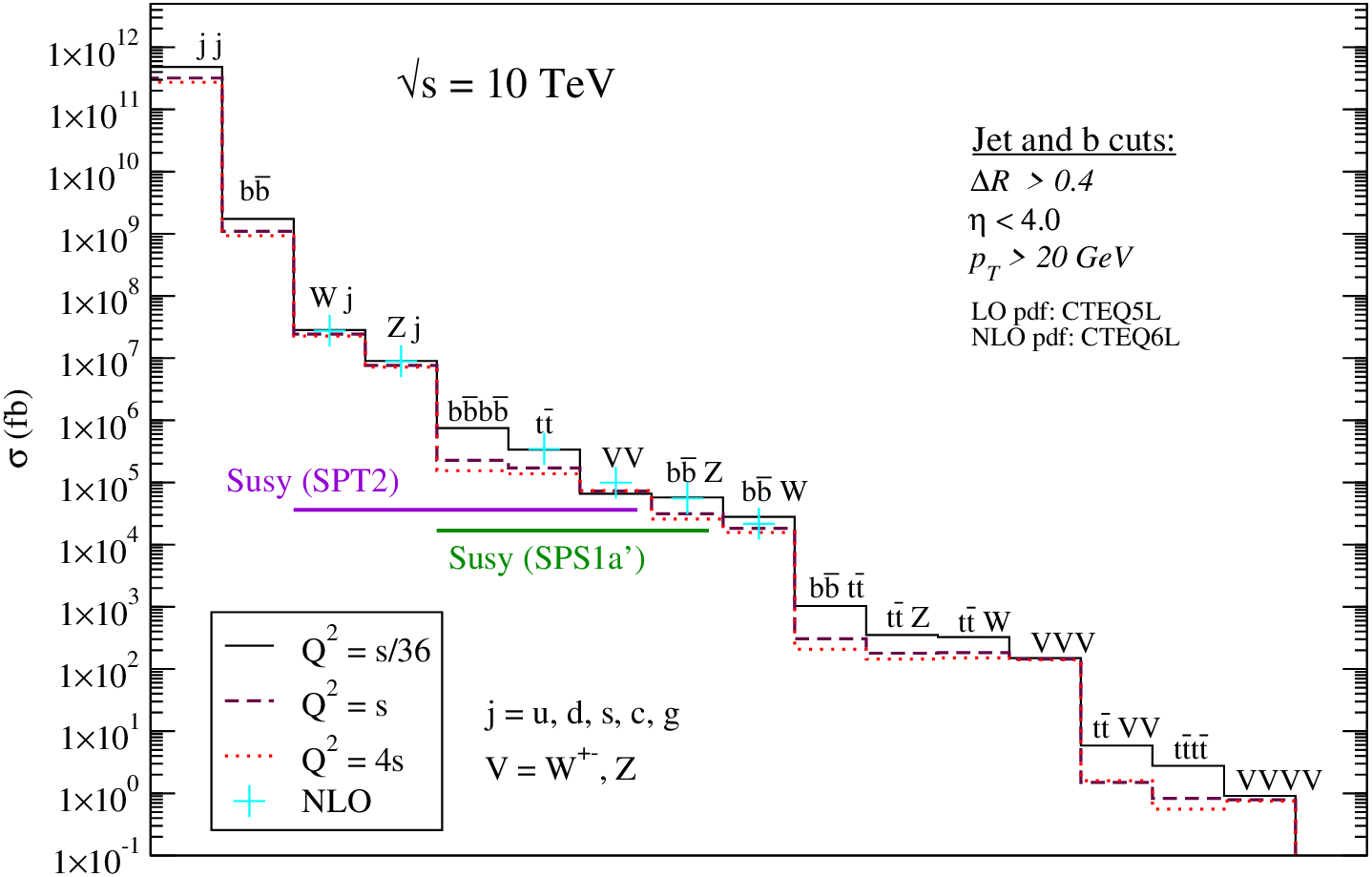}
\caption{ Total cross-sections for several SM backgrounds for $pp$
collisions at $\sqrt{s}=10$ TeV, with three
different choices of renormalization and
factorization scales ($Q$) (taken to be equal) 
shown by the solid ($Q=\sqrt{\shat}/6$),
dashed ($Q=\sqrt{\shat}$) and dotted ($Q=2\sqrt{\shat}$) lines. The NLO
results are shown as blue crosses.  The total cross-section for the
SUSY \sps1ap and SPT2 mSUGRA cases are also shown for comparison
purposes.  }
\label{fig:xsecs10}}
It is well known that at
LO some of these cross-sections strongly depend on the choice of the
renormalization ($\mu_R$) and factorization ($\mu_F$) scales (here we
always take $\mu_F=\mu_R \equiv Q$). To estimate the systematic error
from the scale dependence of the cross sections, we calculated these
for three different scale choices: $Q= 2\sqrt{\shat}$,
$\sqrt{\shat}$ and $\sqrt{\shat}/6$.  As expected, the processes which
exhibit a strong dependence on the scale are the ones with
$\sigma\propto\alpha^n_s$ ($n\ge2$), as seen in Fig. \ref{fig:xsecs10}
and \ref{fig:xsecs14}. In particular, $\sigma(b\bar{b})$,
$\sigma(t\bar{t})$ and $\sigma(b\bar{b}b\bar{b})$ vary by factors of
$1.8$, $2.4$ and $4.6$, respectively.  This scale dependence is
basically the same at $10$ and $14$ TeV, with a small decrease ($\approx
10\%$) for the latter. Using MCFM\cite{mcfm} we computed the NLO total
cross-sections for $t\bar{t}$, $Wj$, $Zj$, $VV$, $b\bar{b}W$ and
$b\bar{b}Z$, with $Q=m_{t}$ for $t\bar{t}$ production, and
$Q^2=m^{2}_{V}+p^{2}_{T}(V)$ for the the processes.
The results are shown in Fig. \ref{fig:xsecs10} and
\ref{fig:xsecs14}. For the dominant backgrounds, namely $Wj$, $Zj$ and
$t\bar{t}$, the NLO results are well approximated by the LO
cross-sections with the scale choice $Q=\sqrt{\shat}/6$. Hence we choose
this scale for all our subsequent background calculations.  We point out that
although our scale choice brings the total LO cross-section closer to
the NLO result, the same is not necessarily true for the different
kinematic distributions used in our analysis. However, to be conservative,
we do not include a K factor for the signal cross-sections.
We also show for comparison the total LO sparticle pair production cross
sections for two mSUGRA points used here as case studies: \bi
\item \sps1ap: $(m_0,\ m_{1/2},\ A_0,\ \tan\beta ,\ sign(\mu )) = (70\ {\rm GeV},\ 250\ {\rm GeV},\ -300\ {\rm GeV},\ 10,\ +)$,
\item SPT2: $(m_0,\ m_{1/2},\ A_0,\ \tan\beta ,\ sign(\mu )) =(450\ {\rm GeV},\ 170\ {\rm GeV},\ 0\ {\rm GeV},\ 45,\ +)$
\ei

The \sps1ap \cite{sps} is a commonly adopted mSUGRA bench-mark point,
while the second point (labeled SPT2 from here on) has a lighter gluino
and slightly heavier squarks than \sps1ap.\footnote{The point \sps1ap has
been selected to yield the correct relic density of neutralino dark matter.
The point SPT2 has much higher neutralino relic density, but is allowed
in scenarios where there exists an axion/axino supermultiplet, in which
case the DM consists of an axion/axino admixture\cite{bbs} rather than
neutralinos.}
Representative sparticle masses for
these cases are shown in Table \ref{table:spectra}.

\begin{table}
\centering
\begin{tabular}{|c|c|c|}
	\hline
mass (GeV) & \sps1ap &  SPT2     \\
	\hline
$\tg$  & 608  & 453 \\
$\tq$  & 555 & 585 \\
$\tst_1$ & 356 & 397 \\
$\tmu_L$ & 191 & 466 \\
$\tmu_R$ & 123 & 455\\
$\tnu_\mu$ & 171 & 458\\
$\ttau_1$& 109 & 348 \\
$m_{\tnu_\tau}$ & 169 &412 \\
$m_{\tw_1}$ & 183 & 114\\
$m_{\tz_1}$ & 98 & 64\\
	\hline
\end{tabular}
\caption[]{Representative sparticle masses for the two mSUGRA case study
points labeled \sps1ap and SPT2 introduced in the text.}
\label{table:spectra}
\end{table}

Though squark and gluino masses are not hugely disparate for the two
points, some aspects of the phenomenology are quite different. In the
\sps1ap case, gluinos decay to squarks (with decays to tops and stops
occurring about 20\% of the time), and $\tq_L \to q'\tw_1$ decays
occurring with a canonical branching fraction close to $2/3$, and
$BR(\tq_R \to q\tz_1) \simeq 1$. For the SPT2 case, squarks mainly decay
to gluinos though $BR(\tq_L \to q'\tw_{1,2})\simeq 0.3$, while gluinos
decay via three body modes. The decay patterns of charginos and
neutralinos, however, differ in an important way between the two cases
because for the \sps1ap case, $\ttau_1$, $\te_R$ and $\tmu_R$ are
significantly lighter than $\tw_1$ and $\tz_2$, while the sneutrinos are
just $\sim 10$~GeV lighter than $\tw_1$ and $\tz_2$.  As a result,
chargino and neutralino decays to stau -- remember that the right
sleptons, being singlets, have no coupling to winos -- are
significantly enhanced, resulting in a softer
spectrum of muons (which
frequently result as secondaries from $\tau$ decays)  in the \sps1ap
case. We will see below that this altered cascade decay pattern has a
significant impact on the early detection of SUSY at the LHC:
although squark and gluino masses are qualitatively similar, the SPT2 point is
accessible at very low integrated luminosities, while detection in the 
\sps1ap case requires considerably larger integrated luminosity.

\FIGURE[bht]{
\includegraphics[width=11cm,clip]{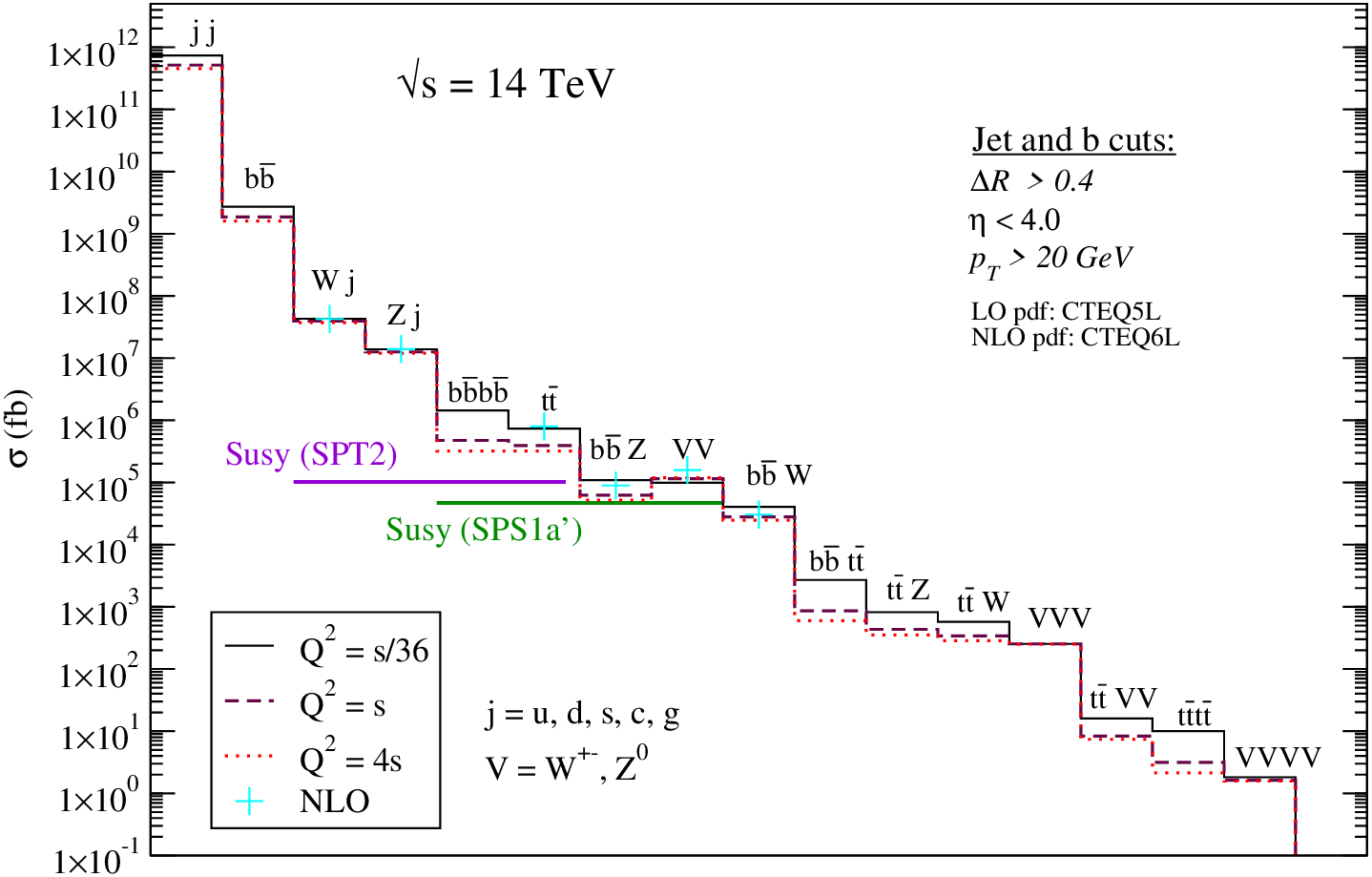}
\caption{Total cross-sections for several SM backgrounds for $pp$
collisions at $\sqrt{s}=14$ TeV, with three
different choices of renormalization and
factorization scales ($Q$) (taken to be equal) 
shown by the solid ($Q=\sqrt{\shat}/6$),
dashed ($Q=\sqrt{\shat}$) and dotted ($Q=2\sqrt{\shat}$) lines. The NLO
results are shown as blue crosses.  The total cross-section for the
SUSY \sps1ap and SPT2 mSUGRA cases are also shown for comparison
purposes. }
\label{fig:xsecs14}}

After verifying that for SUSY searches the most relevant backgrounds are
$t\bar{t}$, $Zj$, $Wj$, $jj$, $t\bar{t}Z$ and $b\bar{b}Z$, we improved
our results by adding multiple jets to these processes.  Using AlpGen
and the MLM matching algorithm\cite{alpgen} (to avoid double counting)
we included in our background simulations the following processes:
$2,3,4$ jets, $t\bar{t}+0,1,2$ jets, $Z+0,1,2,3$ jets, $W+0,1,2,3,4$
jets, $t\bar{t}Z+0,1,2$ jets, $b\bar{b}Z+0,1,2$ jets. In these processes
$Z^{(*)}(\gamma^*)\to l\bar{l}, \nu\bar{\nu}\,(l\bar{l})$ and
$W^{(*)}\to l\nu$.  Since we apply multijet and hard jet $p_T$ cuts in
our analysis (see below) the inclusion of the full matrix element
results for the above processes significantly increases our background
contributions to some of the search channels.

\subsection {Event Simulation} 

For the simulation of the background events we use AlpGen and MadGraph
to compute the hard scattering events and Pythia\cite{pythia} for the
subsequent showering and hadronization.  The signal events were
generated using Isajet 7.78\cite{isajet}.  A toy detector simulation is
then employed with calorimeter cell size
$\Delta\eta\times\Delta\phi=0.05\times 0.05$ and $-5<\eta<5$ . The HCAL
(hadronic calorimetry) energy resolution is taken to be
$80\%/\sqrt{E}+3\%$ for $|\eta|<2.6$ and FCAL (forward calorimetry) is
$100\%/\sqrt{E}+5\%$ for $|\eta|>2.6$, where the two terms are combined
in quadrature. The ECAL (electromagnetic calorimetry) energy resolution
is assumed to be $3\%/\sqrt{E}+0.5\%$. We use the Isajet\cite{isajet}
jet finding algorithm (cone type) to group the hadronic final states
into jets. The jets and isolated lepton definitions are as follow: \bi
\item Jets are required to have $R\equiv\sqrt{\Delta\eta^2+\Delta\phi^2}\leq0.4$ and $E_T(jet)>25$ GeV.
\item Leptons are considered isolated if they have $p_T(l)>5 $ GeV 
with visible activity within a cone of $\Delta R<0.2$ of $\Sigma E_T^{cells}<5$ GeV.

\ei

\subsection{Hadronic resolution and jet energy scale issues for early discovery}

For our analysis of the early SUSY reach, we considered the possibility
that the hadronic energy resolution may not be as good as anticipated,
which could lead to an underestimate of those backgrounds such as
$Z+jets$ that fall steeply with $E_T(j)$. Toward this end, we
re-evaluated the most important backgrounds to dimuon production,
assuming the hadronic energy resolution is only half as good as its
default value above, {\it i.e.}  we take,
\begin{itemize}
  \item Low resolution (Lres): $160\%/\sqrt{E}+3\%$ for $|\eta|<2.6$ and $200\%/\sqrt{E}+5\%$ for $|\eta|>2.6$
\end{itemize}
instead of
\begin{itemize}
  \item Default resolution (Dres): $80\%/\sqrt{E}+3\%$ for $|\eta|<2.6$ and $100\%/\sqrt{E}+5\%$ for $|\eta|>2.6$
\end{itemize}
The results are shown in Table \ref{table:resol}.

\begin{table}
\centering
\begin{tabular}{|c|c|c|c|}
\hline
 & All dimuons (fb)  &  $OS(\mu)$ (fb)    &   $SS(\mu)$ (fb)  \\
\hline
$t\bar{t}+jets$ (Lres) & 60.7 $\pm$ 5.7 & 47.3 $\pm$ 5.4 & 5.1
$\pm$ 1.1\\
$Z+jets$ (Lres) & 80.6 $\pm$ 7.4  & 17.5  $\pm$ 3.4  & 0.0\\
\textbf{Total BG (Lres)} & \textbf{141.3  $\pm$ 9.3 } & \textbf{64.8
$\pm$ 6.4 } & \textbf{5.1  $\pm$ 1.1 }\\
\hline
\textbf{Signal (Lres)}  & \textbf{60.1  $\pm$ 0.6 } & \textbf{39.5
$\pm$ 0.5 } & \textbf{12.0  $\pm$ 0.3 }\\
\hline
$t\bar{t}+jets$ (Dres) & 61.6  $\pm$ 6.0  & 44.6  $\pm$ 5.3  & 3.7
$\pm$ 0.9 \\
$Z+jets$ (Dres) & 66.5  $\pm$ 6.7  & 15.4  $\pm$ 3.2  & 0.0\\
\textbf{Total BG (Dres)} & \textbf{128.0  $\pm$ 9.0 } & \textbf{60.0
$\pm$ 6.2 } & \textbf{3.7  $\pm$ 0.9 }\\
\hline
\textbf{Signal (Dres)} & \textbf{62.7  $\pm$ 0.8 }  & \textbf{41.0
$\pm$ 0.7 } & \textbf{12.2  $\pm$ 0.4 }\\
\hline
\end{tabular}
\caption{Comparison between different calorimeter resolutions for the
$\ge$ 4 jets plus all, OS (with a veto for $m(\mu^+\mu^-)
  \leq 10$~GeV, and 75~GeV $<
m(\mu^+\mu^-) < 105$~GeV) and SS dimuon channels for the dominant SM
background ($Z+jets$ and $t\bar{t}+jets$) and the SUSY \sps1ap
point. The statistical (MC) errors are also shown.}
\label{table:resol}
\end{table}

We see that, with the worse resolution, the $Z \to \mu^+\mu^- + jets$
background 
cross section is indeed increased more than the corresponding cross
section from the signal, or from the top background. However, after the
invariant mass cut to veto $Z$'s the difference is no longer striking. 
We conclude that hadronic calorimetry resolution is unlikely to be an
issue, even for early detection of a signal.  

One may also be concerned about background uncertainty from the jet
energy scale. The $Z(\to \ell^+\ell^-) +j $ cross section is about
100~pb at the LHC, and can be used to establish the jet energy
scale. The variation of the $Z+4j$ cross section due to a 5\%
uncertainty in the jet energy scale is $\pm 20$\% \cite{paigeppc}, which
yields an estimate of the systematic uncertainty for SM background from
this source. Since we will require the signal to background ratio to
exceed 20\% for observability (see Sec.~\ref{sec:early}), uncertainties
in the background from the jet energy
scale also appear to be under control.

\section{Early SUSY discovery: searches at  $\sqrt{s}=10$~TeV with no
  $\eslt$ cuts}
\label{sec:early}

After LHC turn-on in Fall 2009, a period of time will be used for
detector studies and calibration. During this early phase, scheduled for
about eleven months, the LHC will
operate at $\sqrt{s}=10$~TeV, and accumulate about 100-200~pb$^{-1}$ of
integrated luminosity \cite{jenni,paigeppc}. At this time,
the classic SUSY signature of $jets +\eslt$
will almost certainly not be viable 
because of a number of issues related to measurement of
missing transverse energy $\eslt$. 
While weakly interacting neutral particles such as neutrinos or
the lightest neutralinos that escape detection in the experimental
apparatus are the {\it physics} origin of $\eslt$, in practice 
missing transverse
energy
also arises from a variety of other sources, including: \bi
\item energy loss from cracks and un-instrumented regions of the detector,
\item energy loss from dead cells,
\item hot cells in the calorimeter that report an energy deposition
even if there isn't one,
\item mis-measurement in the electromagnetic
calorimeters, hadronic calorimeters or muon detectors and
\item mis-identified cosmic rays in events.
\ei
Thus, in order to have a solid grasp of expected $\eslt$ from
SM background processes, it will be necessary to have detailed knowledge of the
{\it complete detector performance}. Experience at the Tevatron suggests that
this complicated task may well take some time to complete 
at the LHC because many SM processes will have to be scrutinized 
in order to properly calibrate the detector. 
For this reason, SUSY searches using the classic $jets +\eslt$
signature, or for that matter {\it any} signature with $\eslt$
as a crucial requirement, may well take longer than a year 
to yield reliable results.

On the other hand, if sparticles are relatively light -- not far beyond
the reach of Tevatron searches \cite{tevatron} -- then their production
cross sections at the LHC can be huge, and tens of thousands of new physics
events may be produced in the first few months of LHC operation. For
instance, for $m_{\tg}\sim 400$ GeV and heavy squarks, the expected
gluino pair cross sections are in the $10^4$ fb range. If $m_{\tg}\sim
m_{\tq}\sim 400$ GeV, then SUSY production cross sections are even higher: of
order $10^5$ fb!  Thus, with just 0.1 fb$^{-1}$ of integrated
luminosity, we might expect of order $10^3-10^4$ new physics events to
be recorded on tape if the gluino is in the 400 GeV range.  These large
rates provide motivation to re-evaluate search strategies that may be
reliably carried out at the earliest stages of LHC operation at
$\sqrt{s}=10$~TeV. To avoid a complicated analysis of the rate at which
jets fake electrons which will be rather uncertain during
early running, we focus on signals involving only muons and jets, and
where precise determination of the energies does not play a crucial role
in the extraction of the signal over background.  Identification of high
$p_T$ muons, on the other hand, is one of the most straightforward
measurements at LHC, and the ATLAS and CMS detectors are utilizing
cosmic ray muons as a tool for understanding their detectors even before
the LHC turn-on.


In what follows, we define the signal to be observable if
\begin{itemize}
  \item $S \ge max[5\sqrt{B},\ 5,\ 0.2B]$
\end{itemize}
where $S$ and $B$ are the expected number of signal and background
events, respectively. The requirement $S\ge0.2B$ is imposed to avoid
the possibility that a {\it small} signal on top of a {\it large} background
could otherwise be regarded as
statistically
significant, but whose viability would require
the background level to be known with 
exquisite precision
in order to establish a discovery.

\subsection{LHC reach in multi-muon + jets channels without $\eslt$ requirements}
\label{ssec:mm}
%

%
\FIGURE[t!]{ \includegraphics[width=10cm,clip]{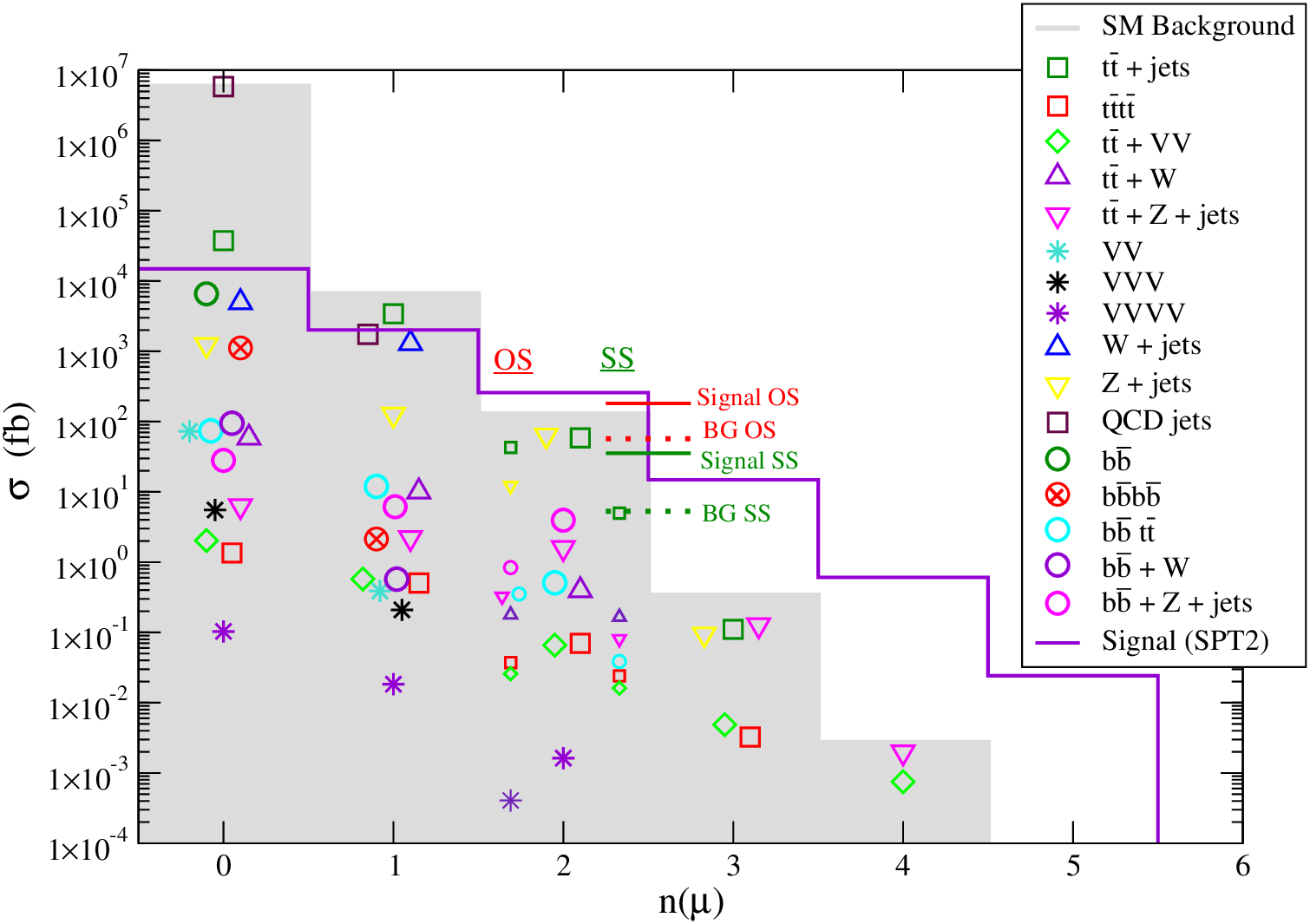}
\caption{Cross sections for various multiplicities of isolated muons in
$n$-muon +$\ge 4$ jet events at the LHC, with $\sqrt{s}=10$ TeV. We show
the signal levels for the SPT2 sample point by the open histogram, along
with corresponding levels for various SM backgrounds. In the $n(\mu)=2$
bin, the left, center and right columns show the background components
for SS, dimuons and OS (with invariant mass cuts), respectively.  }\label{fig:nmu}}

The center-of-mass
energy of 10~TeV is a five-fold increase on the highest collision energies
currently attained and, as just discussed, represents an opportunity
for sparticle searches well beyond the reach of the Fermilab
Tevatron. Motivated by this, we follow up on earlier studies
\cite{bps,bls} and explore the early reach of the LHC in the relatively
straightforward multi-muon plus multi-jet channels where precise energy
measurements are not essential, and complications due to jets faking an
electron are absent.
%
We impose the following basic cuts\footnote{Unless stated otherwise,
these cuts are imposed on all muon plots in what follows.}
\begin{itemize}
  \item Jet cuts: $n(jets)\ge 4$ with 
$E_T(j_1) \ge 100$ GeV, $E_T(j) \ge 50$ GeV and
  $|\eta(j)| \le 3.0$ (jets are ordered $j_1-j_n$, from highest to 
lowest $E_T$)
  \item $S_T \ge 0.2$, where $S_T$ is the transverse sphericity,
  \item Muon cuts: $p_T(\mu) \ge 10$ GeV, $|\eta(\mu)| \le 2.0$, $10$ GeV$ \le m(\mu^+\mu^-) \le 75$ GeV or
$m(\mu^+\mu^-) \ge 105$ GeV (for OS muons only),
\end{itemize}
and plot in Fig.~\ref{fig:nmu} the surviving cross section versus the
muon multiplicity for the SUSY SPT2 point, along with corresponding
\FIGURE[t!]{
\includegraphics[width=10cm,clip]{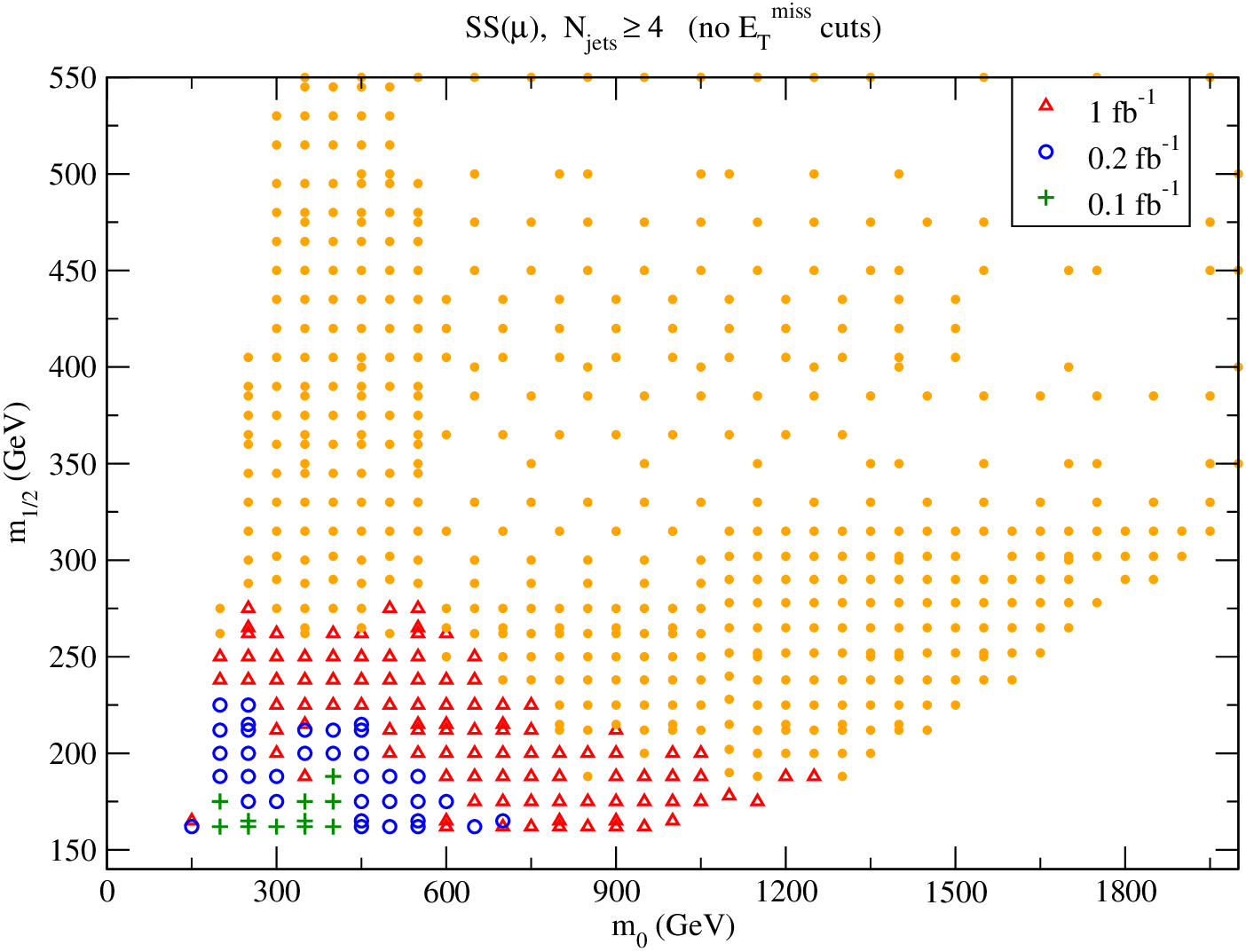}
\caption{ SUSY reach of the LHC at $\sqrt{s}=10$ TeV via SS-dimuon plus
$\ge 4$ jets events with only the basic cuts detailed in the text,
for various integrated luminosities. The fixed mSUGRA parameters
are $A_0=0$, $\tan\beta =45$ and $\mu >0$.
 The solid dots here, and in other
subsequent figures, denote model points where the signal remains
unobservable even for the largest integrated luminosity shown in the
figure.  }\label{fig:SS}}
contributions from a variety of $2\to n$ SM background processes at
$\sqrt{s}=10$ TeV. At low muon multiplicity, signal is well below the
background, which is dominated by QCD multi-jet production for $n_\mu =
0$, and by $t\bar{t}$, $W+j$ and QCD production for $n_\mu = 1$. For
$n_\mu = 2$, we see that the signal and background are already
comparable. We can further divide the dimuon events into the OS and SS
class. For OS dimuons we apply the invariant mass cuts listed above to
avoid the $\gamma^* ,Z\to\mu^+\mu^-$ poles. In this case, the signal
(for this sample point) is seen to exceed the background in both the OS
and SS channels. We also mention that the SS dimuon cross section from 
$W^{\pm} W^{\pm} + dd/uu$ production is negligible: specifically, $\sigma (pp \to
W^+W^+ + X) = 116$~fb, $\sigma(pp\to W^-W^- + X)=46$~fb, and this
contribution to the $\mu^+\mu^++4j$ ($\mu^-\mu^-+4j$) cross section in
Fig.~\ref{fig:nmu} is just 0.015~fb (0.008~fb).
Moving to the $3\mu$ channel, we see that the signal
drops, but the background, which is dominated by $Z$, $t\bar{t}$ and
$t\bar{t}Z$ production, drops even further. The $3\mu\ +\ge 4$ jets
signal is at the 15 fb level, while the corresponding background is
around 0.34 fb: despite the fact that the statistical significance as
well as the $S:B$ ratio are both largest for the trimuon case, an
integrated luminosity in excess of 300~pb$^{-1}$ is necessary to attain
the five-event level that we require for observability.

In Fig. \ref{fig:SS}, we show the reach of the $\sqrt{s}=10$ TeV LHC for
the clean SS dimuon plus $\ge 4$ jets events for various values of
integrated luminosity, using only the basic jet, $S_T$ and muon cuts
mentioned above. We scan over the mSUGRA model parameters $m_0$ and
$m_{1/2}$, with $A_0=0$, $\tan\beta =45$ and $\mu >0$. We see that with
just 0.1~fb$^{-1}$, already values of $m_{\tg}\sim 450$ GeV become
accessible.  As the integrated luminosity is increased to 0.2
(1)~fb$^{-1}$, the reach increases to 550 (650)~GeV.

\FIGURE[t!]{
\includegraphics[width=10cm,clip]{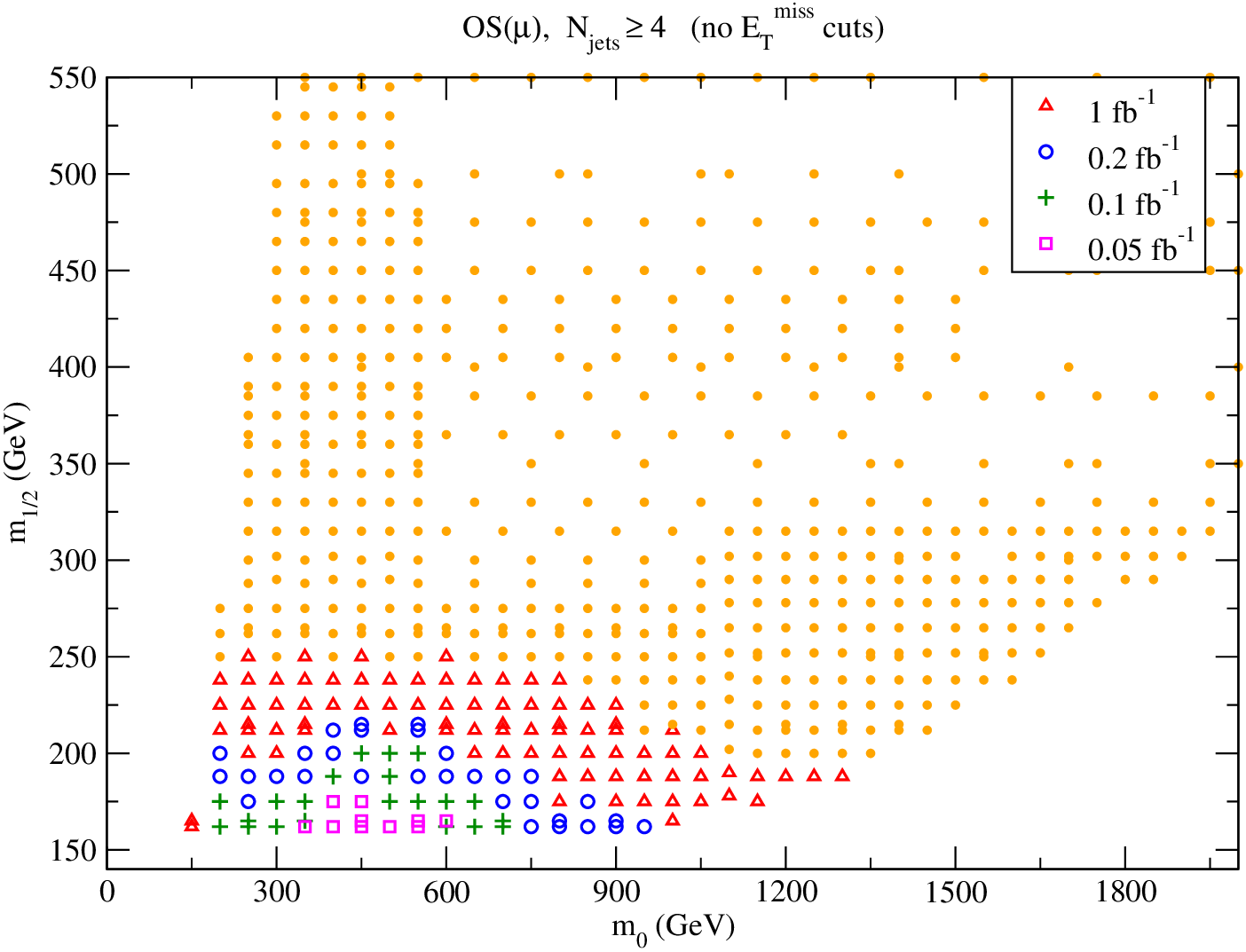}
\caption{ SUSY reach of the LHC at $\sqrt{s}=10$~TeV via OS-dimuon plus $\ge
4$ jets events with only the basic cuts detailed in the text, for
various integrated luminosities.
The fixed mSUGRA parameters
are $A_0=0$, $\tan\beta =45$ and $\mu >0$.
  }\label{fig:OS}}

In Fig. \ref{fig:OS}, we plot the corresponding reach of the
$\sqrt{s}=10$ TeV LHC for OS dimuon plus $\ge 4$ jets events for various
values of integrated luminosity. For low values of $m_{1/2}$, this
signature appears to be even more promising than the SS dimuon channel
since this signal is observable over portions of the parameter space
with $m_{\tg}\alt 450$~GeV with just 50~pb$^{-1}$ of integrated
luminosity!\footnote{We should temper this conclusion with some caution,
since it is contingent upon our background estimate being correct. In
practice, the SM background in the dimuon plus $\ge 4j$ channel will
likely be extracted from the data. Assuming that $t\bar{t}$ events are
the dominant source of this background, we estimate that an integrated
luminosity of 20-30~pb$^{-1}$ will suffice to extract this background
from the measurement of the $t\bar{t}$ cross section with additional
jets.} We see, however, that for larger integrated luminosities (for
which the SS dimuon signal crosses the five-event level), the reach via
the SS channel, which has a larger $S:B$ ratio, exceeds that in the OS
channel. In the OS dimuon channel, the reach in $m_{\tg}$ is 500
(600)~GeV for an integrated luminosity of 0.2 (1)~fb$^{-1}$. 
We should remember that the projections for especially 1~fb$^{-1}$ are
conservative, since it is likely that by the time this is accumulated,
the detectors will  be well enough understood for the $\eslt$ as well as
the electron channels to be useful.


In Fig. \ref{fig:trimu}, we plot the reach of the $\sqrt{s}=10$ TeV LHC
for the trimuon plus $\ge 4$ jets events for various values of
integrated luminosity. Due to the smaller signal cross-section (compared
to the dimuon channels), the trimuon signal remains below observability
for even 200~pb$^{-1}$ of integrated luminosity. However, due to the large
signal to background ratio, even with just 1~fb$^{-1}$, this low-rate but
relatively background-free
channel probes gluino masses up to $\sim 700$~GeV.

\FIGURE[b!]{ \includegraphics[width=10cm,clip]{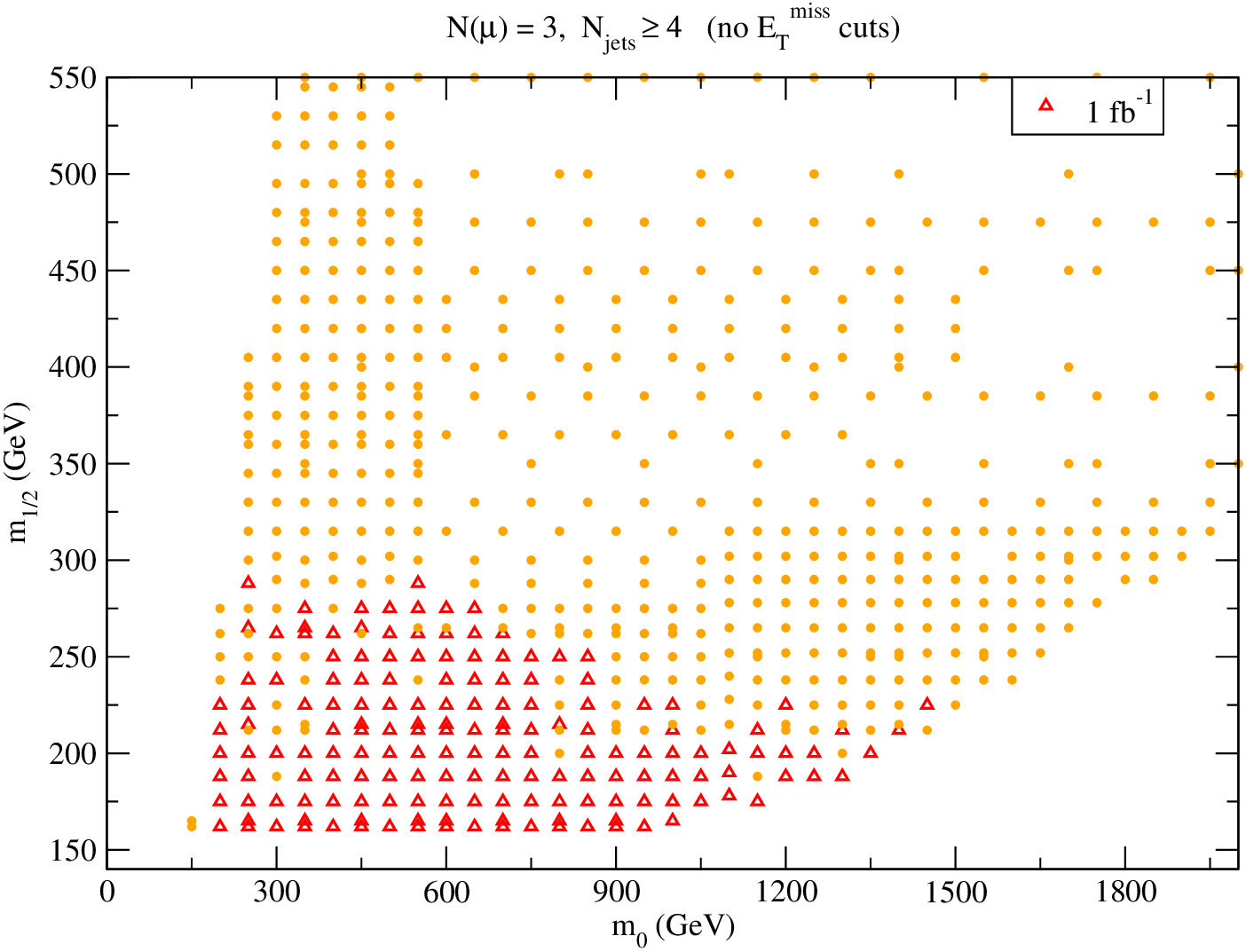}
\caption{ SUSY reach of the LHC at $\sqrt{s}=10$ TeV via trimuon plus
$\ge 4$ jets events with only the basic cuts detailed in the text, for
various integrated luminosities.
The fixed mSUGRA parameters
are $A_0=0$, $\tan\beta =45$ and $\mu >0$.  }\label{fig:trimu}}

\subsection{Characteristics of SUSY multi-muon + jets events}
\label{ssec:amm}

While a discovery of an excess of SS, OS or $3\mu$ plus jets events
would be exciting, it would also be useful to check various aspects of
these multi-muon events to see if they agree with a hypothetical origin
from supersymmetry. This is especially crucial for any discussion of
early SUSY discovery where the signal may initially comprise of just
5-10 events over a very small SM background. With this in mind, we study
various muon (and some jet) distributions: any signal in the ``counting
experiments'' of Sec.~\ref{ssec:mm} will be that much more convincing if
these events have the expected characteristics discussed below. As the
accumulated integrated luminosity grows, these same distributions (with
electrons combined with muons) will, of course, provide more precise
information about sparticle masses, which will help to zero in on the
underlying model. We show these distributions for the two SUSY cases
\sps1ap and SPT2 introduced above as well as for the SM background,
beginning with OS dimuon channel which has the potential for the
earliest discovery of SUSY.

\subsubsection{Opposite sign dimuon + jets events}
\label{sssec:os}

We begin by illustrating in Fig. \ref{fig:phiOS} the distribution of the
transverse plane opening angle $\Delta\phi (\mu^+\mu^- )$ between the
muons on $OS$ dimuon $+\ge 4$ jets events from the \sps1ap (dashed),
the SPT2 (solid) SUSY cases  and from the SM background (shaded).
We see that for the SPT2 point where a large fraction of the muons
originate from high $p_T$ $\tz_2\ (\to \mu^+\mu^-\tz_1)$ 
produced in gluino and squark
cascade decays; this distribution peaks at small angles. The distribution tail
comes from dimuons originating from cascade-decay-produced charginos
and extends out to $\Delta\phi(\mu^+\mu^-)\sim \pi$.
\FIGURE[tb!]{
\includegraphics[width=9.0cm,clip]{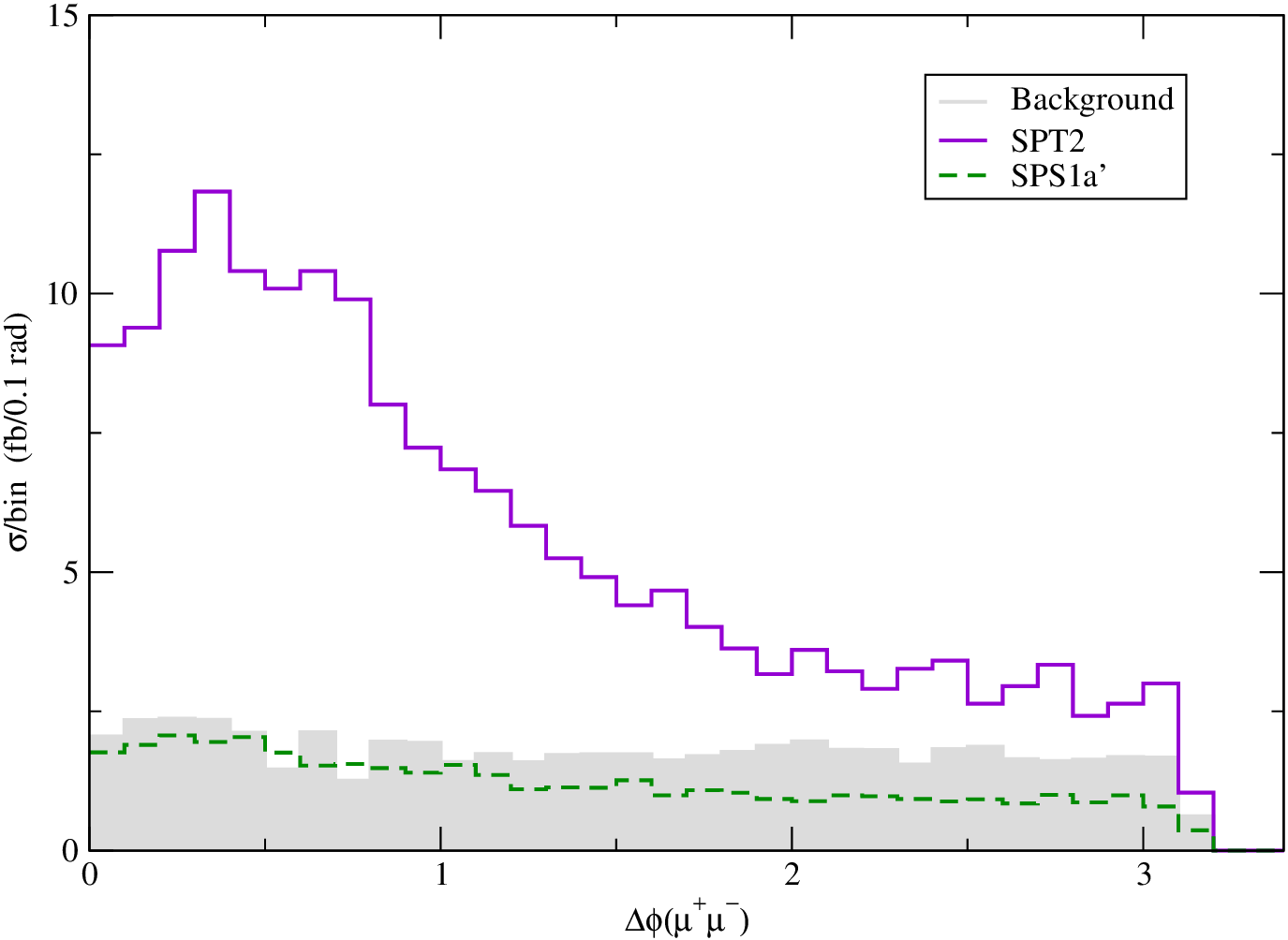}
\caption{
$\Delta\phi (\mu^+\mu^- )$ distribution from OS dimuon $+\ge 4$ jets 
events for \sps1ap (dashed) and SPT2 (solid) cases, and also for
SM backgrounds (shaded). We make no requirement on $\eslt$.
}\label{fig:phiOS}}
For the \sps1ap point, the stau is light,
and so a much smaller fraction of muons come from direct decays of
$\tz_2$, and the corresponding distribution is much flatter. The
difference between the two signal distributions is a reflection of the
different origins of the muons in the two cases.   
The SM background is nearly flat, but also
with a slight peak at low values of $\Delta\phi$. 

\FIGURE[t]{
\includegraphics[width=10cm,clip]{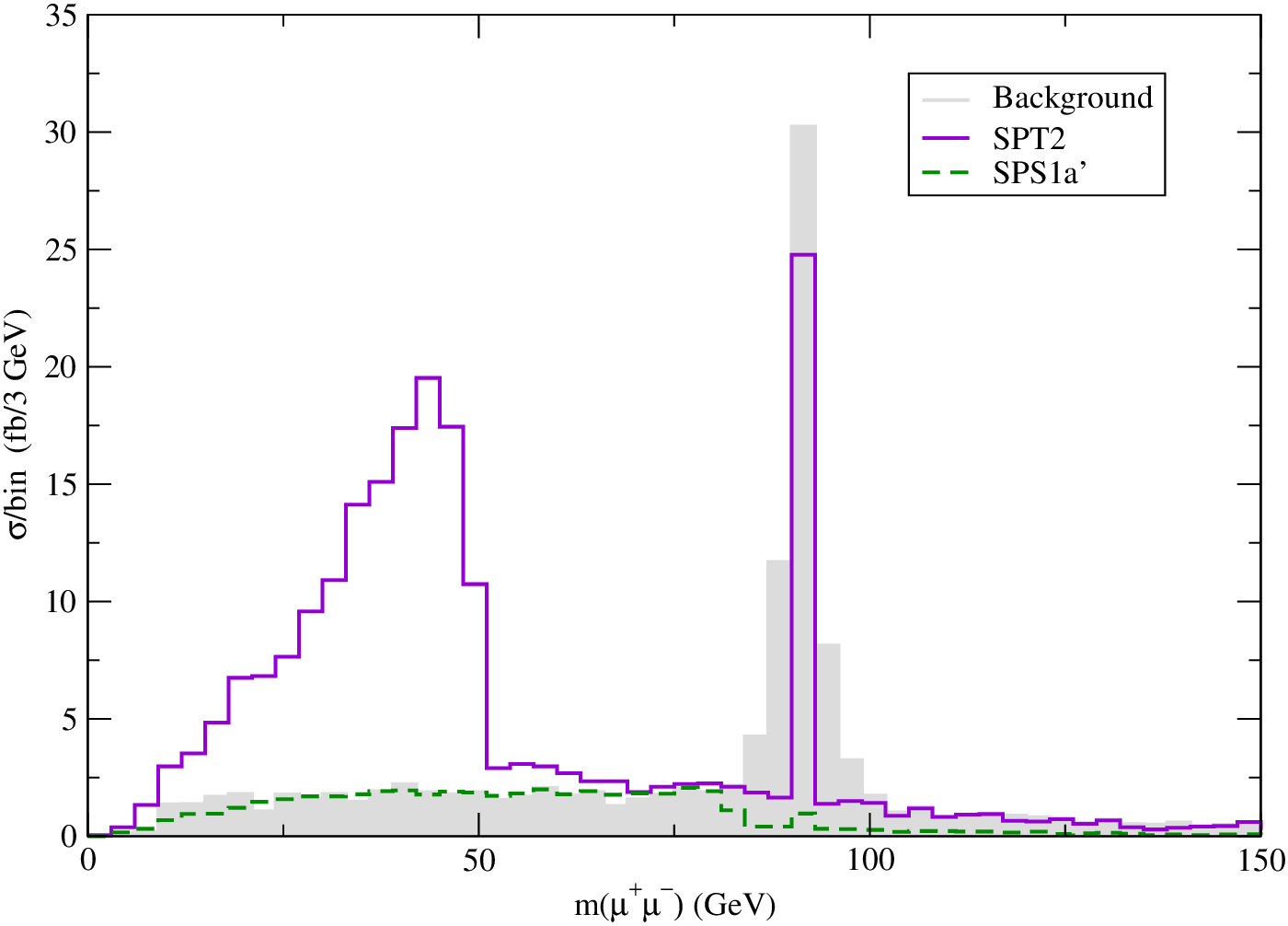}
\caption{
OS dimuon invariant mass distribution from OS dimuon $+\ge 4$ jets 
events for \sps1ap (dashed) and SPT2 (solid) cases, and also for
SM backgrounds (shaded). In this plot only we do not apply
the invariant mass cuts for OS dimuons.
We make no requirement on $\eslt$.
}\label{fig:mOS}}

The distribution of the invariant dimuon mass $m(\mu^+\mu^- )$ in OS
dimuon $+\ge 4j$ events shown in Fig. \ref{fig:mOS} is even more
distinctive. In this case, the SPT2 signal distribution shows the
distinctive kinematic mass edge\cite{mlledge} at 
$m(\mu^+\mu^-)= m_{\tz_2}-m_{\tz_1} = 50.6$ GeV.
In this case, most of the signal dimuons would be expected to cluster
just below 50~GeV, strengthening the case for the SUSY origin of a
signal in the earliest data set.  For the \sps1ap case (which, we
emphasize, is somewhat atypical), the mass edge from $\tz_2 \to \tmu^\pm\mu^\mp
\to \mu^+\mu^-\tz_1$ decays (which have a branching fraction of just 2.4\%)
at $\sim 82$~GeV is considerably less distinctive. Moreover, this edge
merges right into the Z peak, and so will not be measurable.
Furthermore, notice that this distribution is smeared out to low mass
values because both $\tz_2$ and $\tw_1$ decay dominantly to third
generation sleptons, and the muon is frequently produced as a secondary
from tau decays. Except for the $Z$-peak, the SM background is
featureless.
%

Finally, in Fig. \ref{fig:meffOS} we show the $\sum E_T(jet)$
distribution for events with OS dimuons $+\ge 4$ jets events: the large
energy release occurring in sparticle pair production and decay provides
a harder distribution than that expected from SM background. We see that
in both cases it should be possible to pick out the signal over the
background using only the measured transverse energies of the jets,
though for the \sps1ap case, with the smaller cross section, a somewhat
larger integrated luminosity will be required.

\FIGURE[hb]{
\includegraphics[width=10cm,clip]{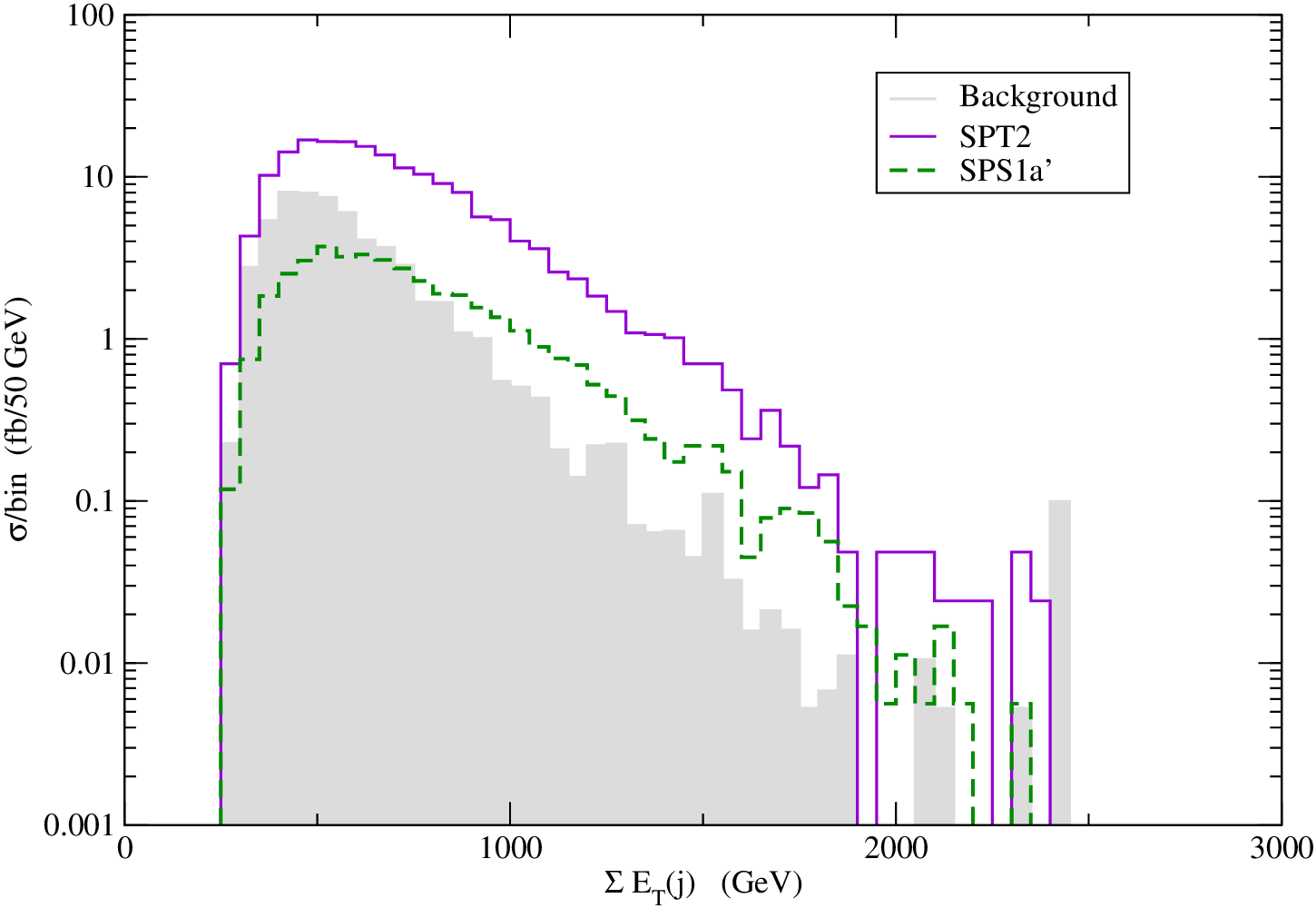}
\caption{
Distribution of $\sum E_T(j)$ for OS dimuon $+\ge 4$ jets events
from signal cases \sps1ap (dashed)  and SPT2 (solid), and from SM
backgrounds. We make no requirement on $\eslt$.
}\label{fig:meffOS}}

\newpage
\subsubsection{Same sign dimuon + jets events}
\label{sssec:ss}

As mentioned in Sec. \ref{ssec:mm}, the SS dimuon channel requires larger
integrated luminosity (due to its reduced rate), but provides a much cleaner
signal. To get an idea of the overall $p_T(\mu)$ distribution in this channel,
we show in Fig. \ref{fig:ptmu} this distribution for the hard ($\mu_1$)
and soft ($\mu_2$) muons from SS dimuon events without any jet cuts for
the signal points \sps1ap and SPT2 and for the background, which mainly 
comes from $t\bar{t}$ production.  While the highest $p_T$ muon
from $t\bar{t}$ production comes from $t\to bW$ followed by $W\to\mu\nu_\mu$ decay,
and is quite hard, the lower $p_T$ muon must come from
$b\to c\mu\nu_\mu$ decay, and hence is necessarily soft, since there
is much less energy release in $b$ decays. Thus, while the signal emerges from
the background only for $p_T(\mu_1)\agt 100-125$~GeV, the soft muon from
the signal -- likely arising from some heavy sparticle decay -- has a much
harder distribution than the corresponding background muon. For this
reason, we require $p_T(\mu_2 )>10$ GeV, even though it might be
possible for LHC detectors to go even lower in muon transverse
momentum. We see again that the \sps1ap case is not the norm in that
especially the lower energy
signal muon frequently arises from tau decay, and so is also soft.
\FIGURE[t]{
\includegraphics[width=9cm,clip]{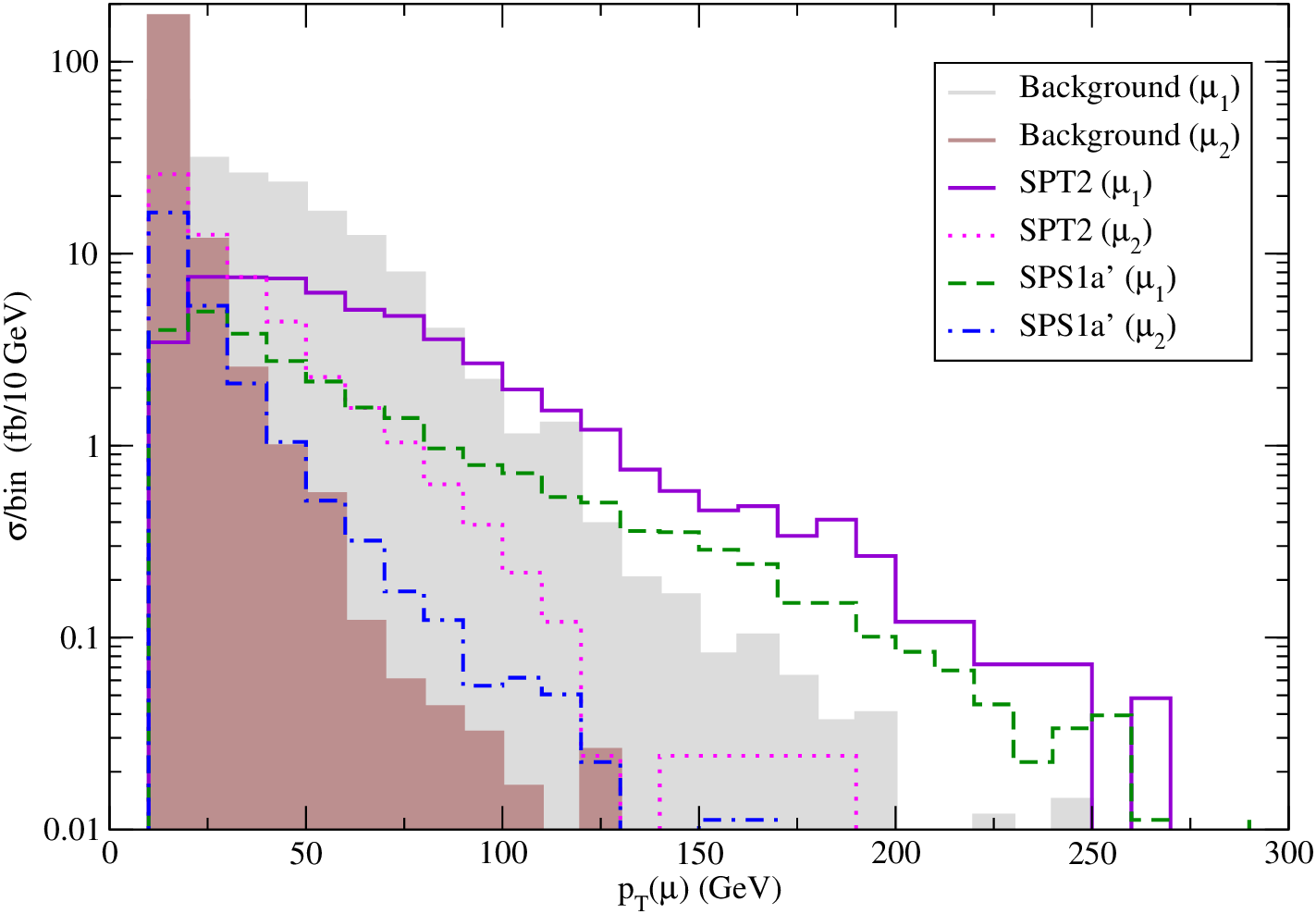}
\caption{ $p_T$ distribution for the harder ($\mu_1$) and softer
($\mu_2$) muons in SS dimuon events for the SUSY \sps1ap and SPT2 mSUGRA cases, along with
corresponding SM background distributions. We make no requirement on
number of jets or $\eslt$.  }\label{fig:ptmu}}

Next, we show in Fig. \ref{fig:phiSS} the distribution of the transverse
plane opening angle between the two muons in SS events with $\ge 4$
jets, again for both \sps1ap and SPT2 signals and for the SM
background. This distribution differs sharply from the corresponding
distribution for OS dimuon events shown in Fig.~\ref{fig:phiOS} in that
the signal is peaked near $\Delta\phi\sim \pi$. This shape  is merely a
reflection of the fact that in the SS case the two muons typically
originate from {\it different} primary particles in the SUSY $2\to 2$
($\tg\tg$, $\tg\tq$, or $\tq\tq$) production subprocess, in contrast to OS dimuons
from neutralino decays.
%
\FIGURE[bt!]{
\includegraphics[width=9cm,clip]{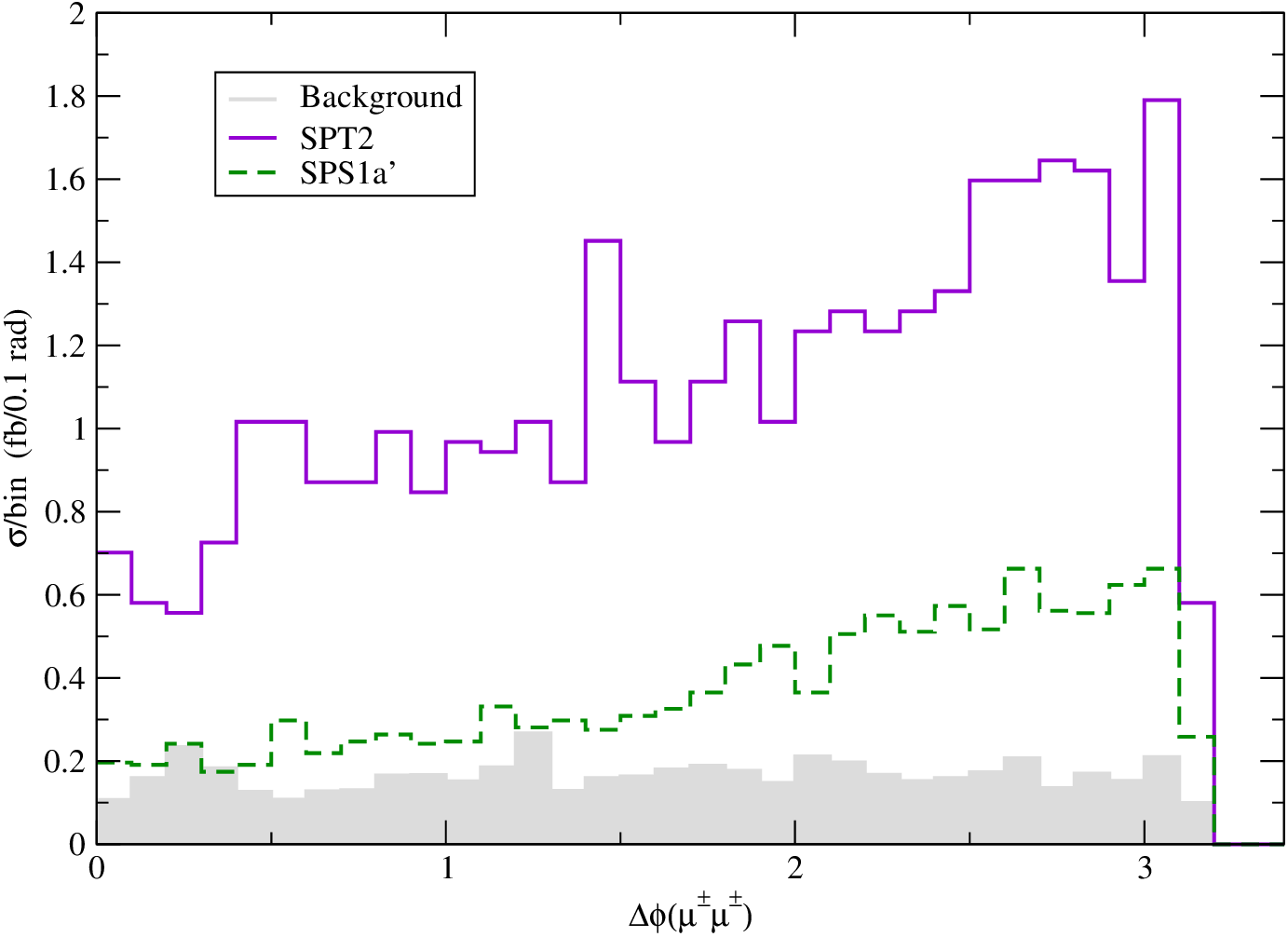}
\caption{Distribution of the transverse plane opening angle between the
  muons,
$\Delta\phi(\mu^\pm\mu^\pm)$, from SS dimuon $+\ge 4j$ events for the
SUSY cases \sps1ap and SPT2, along with the corresponding distribution
from SM sources. We make no restriction on $\eslt$.}\label{fig:phiSS}}
The background distribution is
nearly flat in $\Delta\phi (\mu^\pm\mu^\pm )$.

We have checked that the shape of the distribution of $\sum E_T(j)$ in
SS dimuons + $\ge 4j$ events is qualitatively similar to that for OS
dimuon events shown in Fig.~\ref{fig:meffOS}, and so we do not show it
here.

%
%

%
%

\subsubsection{Trimuon + jets events}
\label{sssec:3m}

Because it has an even smaller background, the trimuon channel -- at a
high enough integrated luminosity -- could
potentially become the best muon channel for SUSY searches without
$\eslt$ cuts.  In Fig. \ref{fig:m3l}{\it
a}, we show the trimuon invariant mass distribution from $3\mu +\ge 4$
jets events for the SUSY points \sps1ap and SPT2, along with that from
the SM background at $\sqrt{s}=10$ TeV.  As expected, the signal
distribution is relatively featureless, since at least two of the muons
originate in different parent particles. The trimuon cross section is
small -- 14.8~fb for SPT2 point, and 3.2~fb for the \sps1ap case -- so
that even with 1~fb$^{-1}$ only the former case leads to an observable
signal in Fig.~\ref{fig:trimu}.
\FIGURE[b]{
\includegraphics[width=6cm,clip]{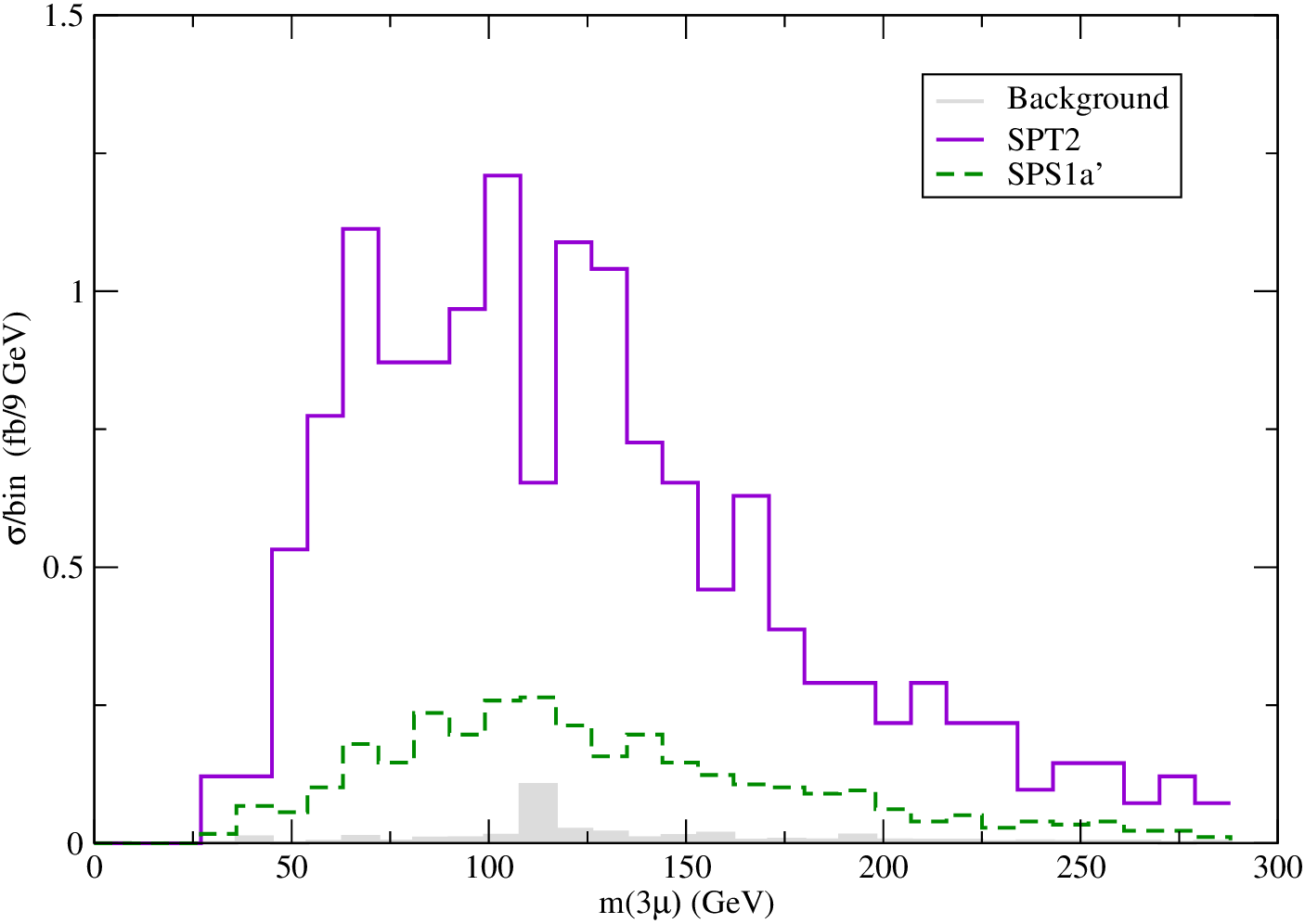}\hspace{.5cm}
\includegraphics[width=6cm,clip]{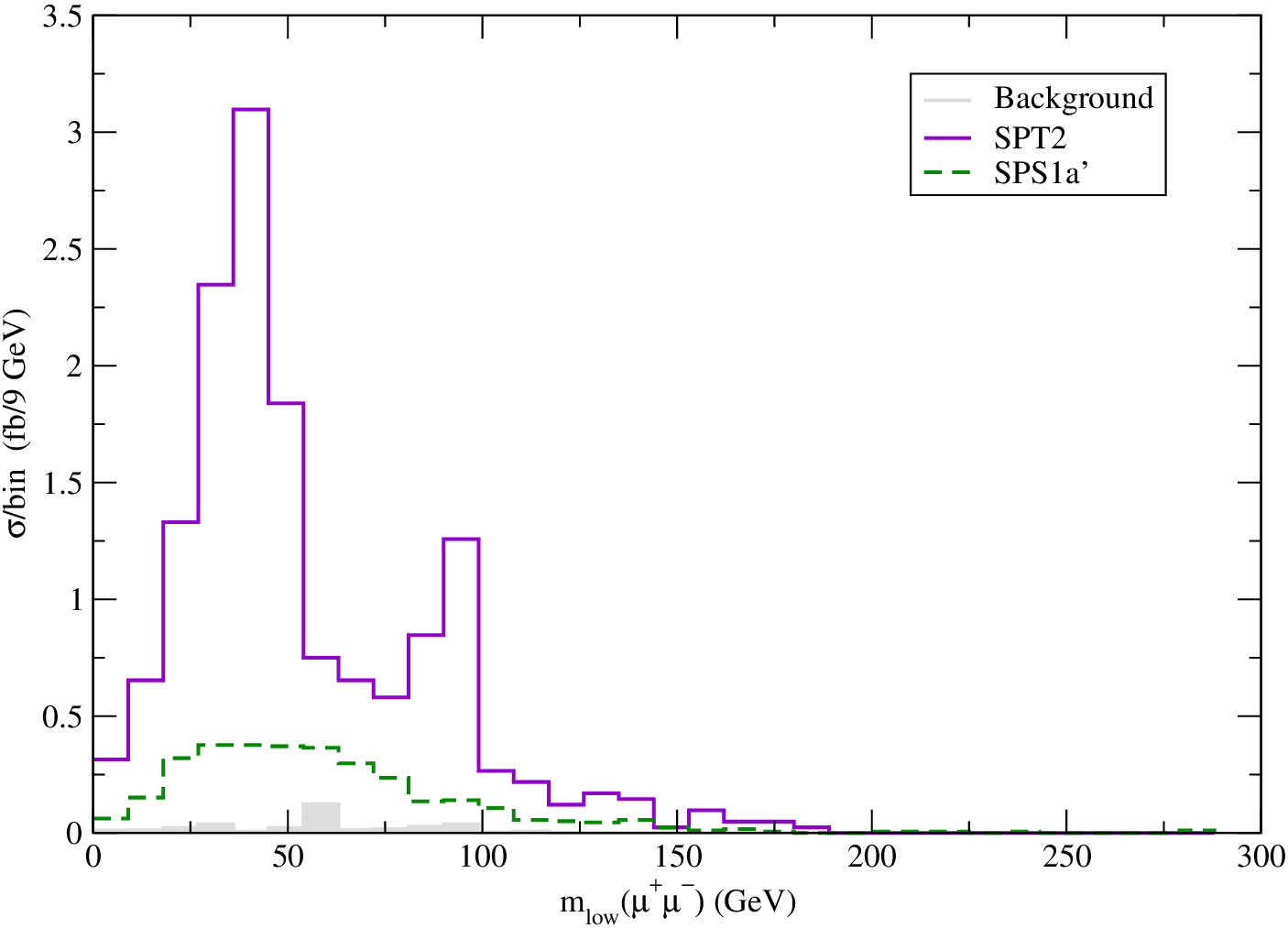}
\caption{The distribution of ({\it a})~trimuon mass, and ({\it b})~the
  smaller of the two OS dimuon invariant masses, from trimuon $+\ge 4$ jets events
  at the 10~TeV LHC for SUSY points \sps1ap and SPT2, along with SM
  backgrounds.  }\label{fig:m3l}}

The noteworthy thing, however, is that though the signal is very small,
it is almost background-free, even with just the basic cuts. Thus this channel
offers prospects for a striking confirmation of new physics 
(presumably first discovered in the OS dimuon channel). In most SUSY models we
expect that $m_{\tw_1}$ is comparable to $m_{\tz_2}$, so that if
$\tz_2$ can be produced in gluino and squark cascade decays, the
chargino can usually also be produced via these decay cascades. Thus a
subsample of trimuon events is likely to include an OS muon pair from
$\tz_2$ decays. With this in mind, we show the distribution of the {\it
smaller} of the two OS dimuon invariant masses in these trilepton +$\ge 4j$ events
in frame {\it b}) of this figure. For both cases, we see a mass edge at
essentially the same location as in Fig.~\ref{fig:mOS}, and further,
that below the mass edge, the two distributions are very similar,
reflecting the common parentage of the dimuons in the two cases. In
favorable cases such as SPT2, the trimuon signal could thus make a
strong case for the SUSY origin of a signal first seen with just $\sim
50-200$~pb$^{-1}$ of the LHC data. For the \sps1ap point, the trimuon
signal offers the possibility of determining the mass edge (obscured by
the $Z$ peak in the OS dilepton case), though an integrated luminosity
of $\sim 10$~fb$^{-1}$ may be required (by which time it is likely that
electron events will also be possible to include in the trilepton
sample).\footnote{Events beyond the mass edge arise, for instance, from
$\tg \to t\tst_1$ decays,  as well as from $Z$s produced in $\tz_3$ decays in
the SPT2 case.}

Finally, we remark that the $\sum E_T(j)$ distribution for $3\mu\ +\ge 4$
jets events is again harder than that for the SM background; since its
features are again much the same as in Fig. \ref{fig:meffOS}, we do not
show it here.

%
%

\subsubsection{Early LHC reach via multimuons plus jets: optimized cuts}
\label{sssec:opt_noetm}

We have seen from the various reach plots presented in Sec. \ref{ssec:mm} 
that even with very basic cuts, and just $0.1-1$~fb$^{-1}$ of
integrated luminosity, LHC experiments will probe gluino and squark
masses well beyond the reach of the Tevatron. These 
cuts have not, however, been optimized over different 
regions of parameter space, and so will not yield the 
maximal reach for a given value of integrated luminosity. In this section,
we implement a large grid of potential cut values, and then select 
the grid (set of cuts) value which optimizes the reach. We recognize
that this may be only of academic interest in that by the time such an
analysis is actually carried out, LHC detectors may be understood well
enough to allow the inclusion of electrons as well as $\eslt$ in the
analysis. We nevertheless felt that it would be worthwhile to explore
just how much information can be gleaned from the data, in  case  
circumstances make this necessary.

We begin with a set of pre-cuts:
\begin{itemize}
\item transverse sphericity $S_T\ge 0.2$,
\item for isolated muons: $p_T(\mu )\ge 10$ GeV and  $|\eta (\mu )|<2.0$,
\item $n(jets)\ge 2$ with $E_T(j)\ge 50$ GeV and $|\eta (j)|<3.0$,
\item for OS dimuons: 10 GeV$\le m(\mu^+\mu^- )\le 75$ GeV or $m(\mu^+\mu^- )> 105$ GeV
\end{itemize}

Then, for a given point in mSUGRA parameter space with of order 50,000 events
generated, we find the optimal set of cuts to maximize $S/\sqrt{B+S}$, 
using:
\begin{itemize}
\item $n(jets)\ge 2,\ 3,\ 4,\ 5,\ 6,\ 7$
\item $E_T(j_1)\ge 50,\ 80-340$ GeV (in steps of 20 GeV), $400-1000$ GeV
(in steps of 100 GeV),
\item $E_T(j_2)\ge 50,\ 55-205$ GeV (in steps of 15 GeV), $300,\ 400,\ 500$ GeV,
\item number of isolated muons $n(\mu )= 2,\ 3,\ 4,\ 5,\ 6$
\end{itemize}
We do not implement any $\eslt$ requirement.
We take $\sqrt{s}=10$ TeV, and adopt standard parameters $A_0=0$,
$\tan\beta=45$ and $\mu >0$.  The results of our optimized cuts analysis
is shown in Fig. \ref{fig:sug10} for various integrated luminosity
choices.  We note the following:
\bi
\item for low integrated luminosities ($0.05$ fb$^{-1}$ and $0.1$
fb$^{-1}$) the optimal cuts are $n(jets)\ge 4,5$ with $E_T(j_1)\sim100$
GeV and $E_T(j_2)\sim70$ GeV in the total dimuon channel. 

\item the same is also true for $0.2$ fb$^{-1}$ and $1$ fb$^{-1}$,
except for $m_0\gtrsim 1200$ GeV and/or $m_{1/2}\gtrsim320$ GeV, where
the $n(jets)\ge 6,7$ channels become more favorable. Moreover the
preferred jet $E_T$ cuts are always below 200 GeV for $0.05,\ 0.1$ and
$0.2$ fb$^{-1}$ and below 300 GeV for $1$ fb$^{-1}$.  \ei We caution the
reader that the projected reaches will be sensitive to the uncertainty
in our estimate of the background, especially for high jet
multiplicities. It is, however, worth bearing in mind that even after
optimization, the bulk of the parameter space is probed with $n_j\ge
4,5$ and $n_\mu=2$.\footnote{This has the added advantage that even the
``trials factor'' is greatly reduced.} The optimized low
$m_0$ reach from Fig. \ref{fig:sug10} extends up to $m_{1/2}$ values of
225, 275 and 325 GeV for integrated luminosity values of 0.1, 0.2 and 1
fb$^{-1}$ respectively.  This corresponds to a reach in terms of
$m_{\tg}$ of about 600, 700 and 800 GeV, respectively.

\FIGURE[t!]{
\includegraphics[width=10cm,clip]{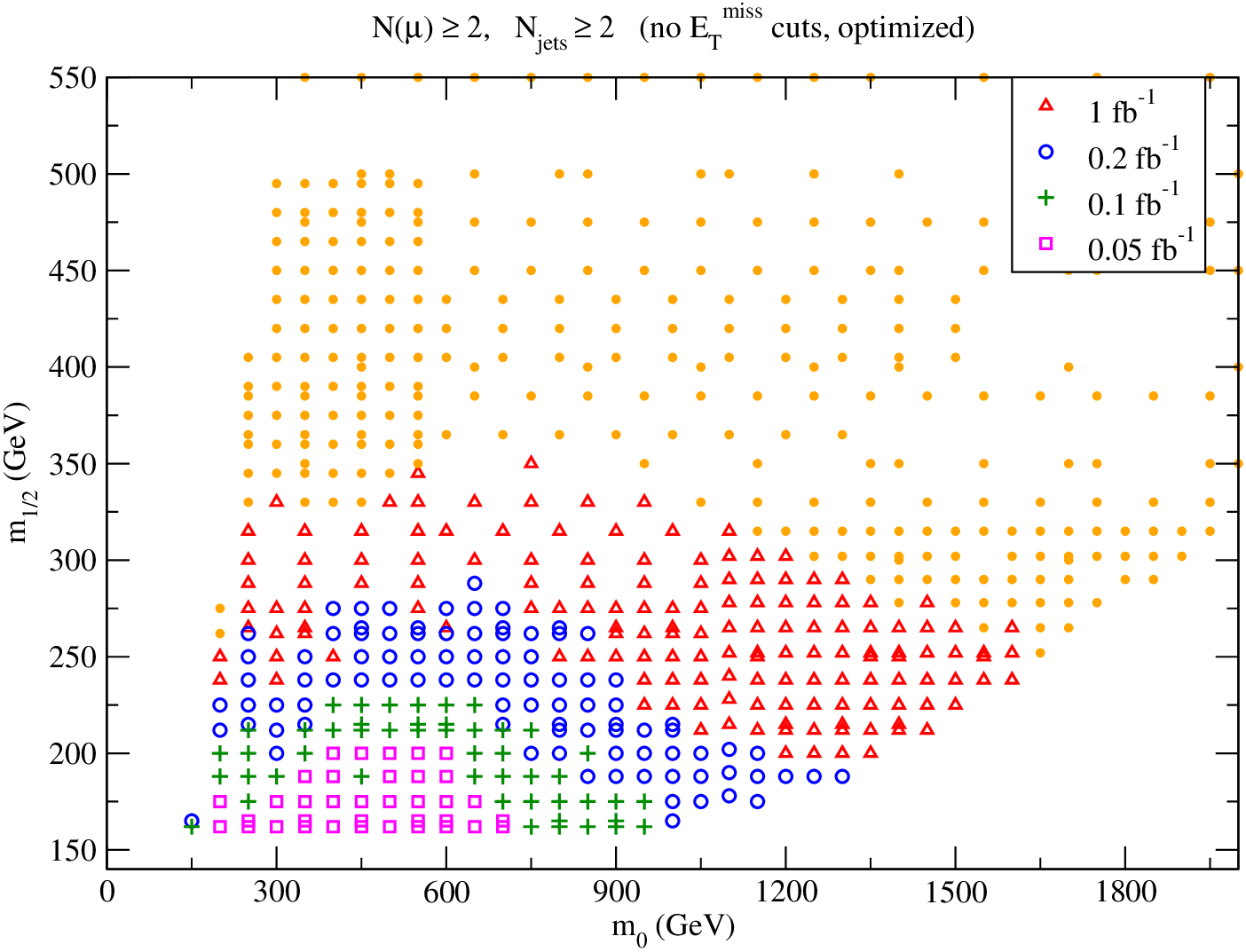}
\caption{ Reach of LHC for mSUGRA at $\sqrt{s}=10$ TeV for multi-muon
$+jets$ events using optimized cuts discussed in the text, but without
any $\eslt$ requirement on the signal.
The fixed mSUGRA parameters
are $A_0=0$, $\tan\beta =45$ and $\mu >0$.  }\label{fig:sug10}}

With accumulation of integrated luminosity, the detectors will rapidly
become better understood and reliable electron identification will be
possible.  It will then be possible to use different flavor, OS dilepton
distributions to statistically subtract chargino and $W$ contributions
from the same flavor, OS dilepton signal and sharpen up the dilepton
mass edge first obtained in Fig.~\ref{fig:mOS}.  Using the same cuts
(except that we now include electrons) as in Fig.~\ref{fig:phiOS} and
\ref{fig:mOS}, we plot the distribution of 
$\Delta\phi(e^{\pm}\mu^{\mp})$ in Fig.~\ref{fig:phiOF}, and  the
``subtracted'' like-flavor, OS dilepton mass distribution, 
\begin{equation}
\frac{d\sigma}{dm}(e^+e^- +\mu^+\mu^- -e^+\mu^- -e^-\mu^+)=
\frac{d\sigma(e^+e^-)}{dm_{ee}}+\frac{d\sigma(\mu^+\mu^-)}{dm_{\mu\mu}}
-\frac{d\sigma(e^+\mu^-)}{dm_{\mu e}}-\frac{d\sigma(e^-\mu^+)}{m_{e\mu}},
\label{eq:asym} \end{equation}
in Fig.~\ref{fig:Asymm}.
\FIGURE[t]{
\includegraphics[width=9.2cm,clip]{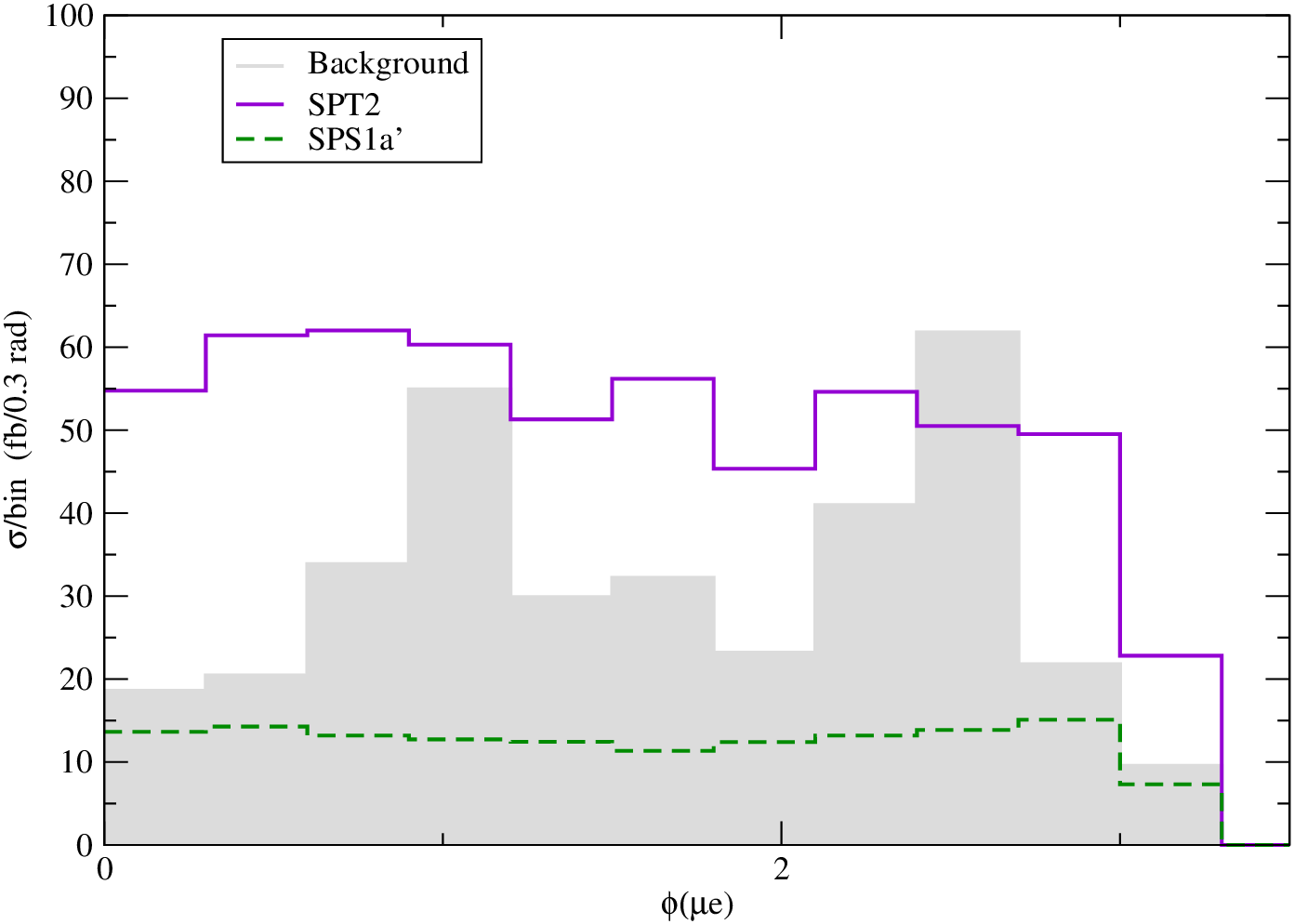}
\caption[]{ $\Delta\phi(e^{\pm}\mu^{\mp})$ for the mSUGRA points \sps1ap
and SPT2 at $\sqrt{s}=10$, along with SM backgrounds. The lepton and jet
cuts are as in Fig.~\ref{fig:phiOS}, and there is no requirement
on $\eslt$.}
\label{fig:phiOF}}
\FIGURE[htb]{
\includegraphics[width=9.2cm,clip]{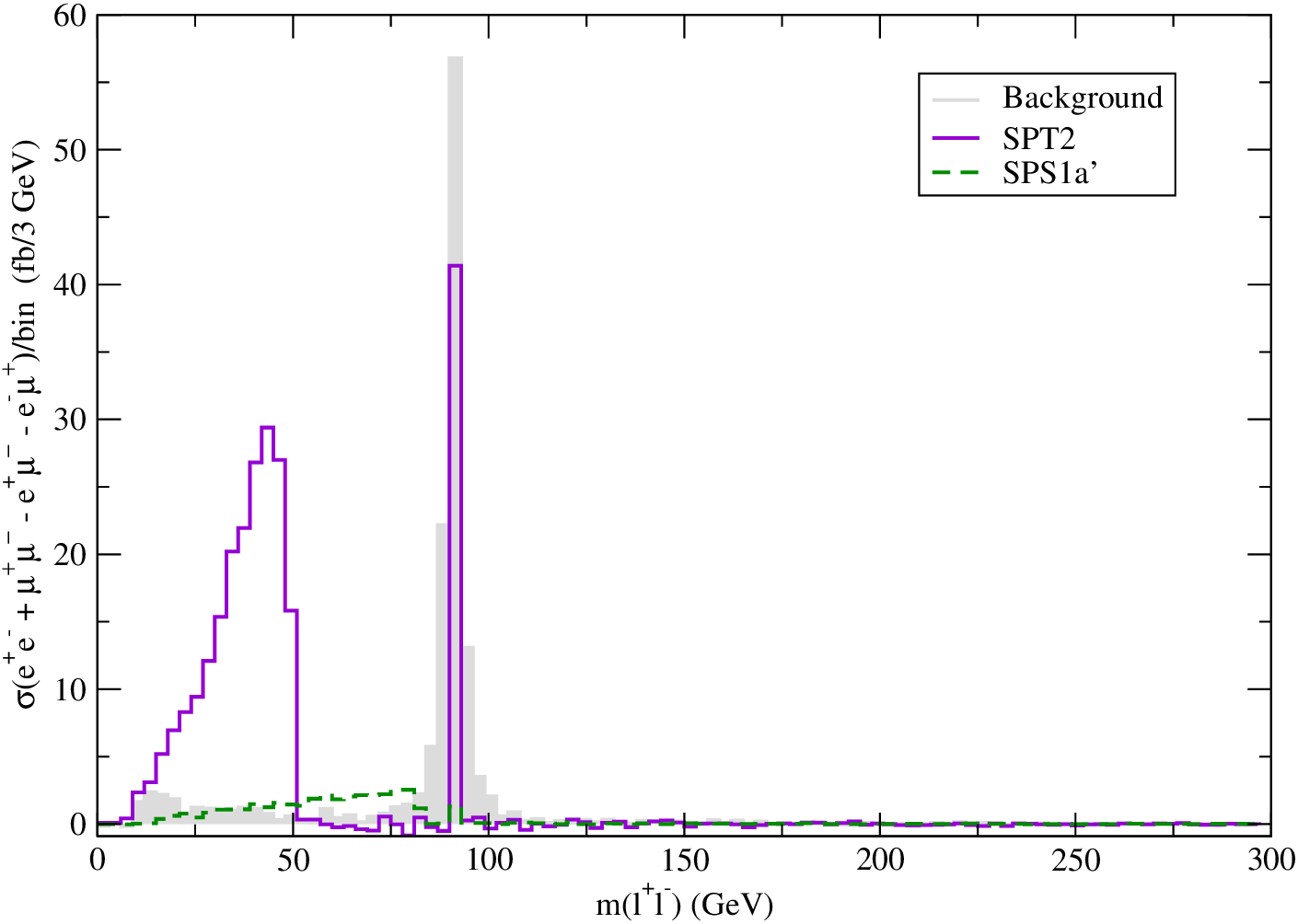}
\caption[]{ The subtracted like-flavor, OS dilepton mass distribution
for the mSUGRA points \sps1ap and SPT2 at $\sqrt{s}=10$, along with SM
backgrounds. The lepton and jet cuts are the same as in
Fig.~\ref{fig:mOS}, and there is no requirement on $\eslt$.}
\label{fig:Asymm}}
\newpage
\noindent We see that the azimuthal angular distribution of
$\Delta\phi(\mu^\pm e^\mp)$ is rather flat, as may be expected since 
the leptons arise from various decay chains including $\tg \to tb\tw_1$
(SPT2 case) and $\tg \to t\tst_1$ (\sps1ap case) which can give
$e^\pm\mu^\mp$ pairs from the decay of the same gluino in addition to
$e^\pm\mu^\mp$ pairs where the electron and muon each originate in
a different gluino (or squark) parent (as in the SS dilepton case).
As anticipated, the dilepton mass edges become significantly sharper
upon flavor-subtraction, though the $Z$ peak continues to obscure the
edge in the \sps1ap case. Of course, this subtraction procedure ought to
work equally well for the trilepton signal shown in
Fig.~\ref{fig:m3l}{\it b}, with the understanding that the dilepton
pairs in (\ref{eq:asym}) now refer to the OS dilepton pair in trilepton
events with the smaller of the two masses.

\subsection{Early LHC reach in dijet channel at $\sqrt{s}=10$ TeV}

Recently, Randall and Tucker-Smith (RT-S)\cite{rts} have proposed a search for
SUSY in the dijet channel, also without using $\eslt$.
While SUSY dijet + $\eslt$ searches have been proposed for a long time,
RT-S emphasized that the search can be made without recourse to an
$\eslt$ cut. RT-S propose selecting events with exactly two jets with
$E_T>50$ GeV and no isolated leptons. This probes the small $m_0$
region of the mSUGRA space because $\tq_R\tq_R$ pair production
naturally leads to this event topology, since for $m_{\tq}< m_{\tg}$,
$\tq_R$ mainly decays via $\tq_R\to q\tz_1$.
RT-S then examine distributions of {\it a})~$\alpha\equiv
E_T(j_2)/m(j_1j_2)$, {\it b})~$\Delta\phi (j_1j_2)$ (the dijet
transverse plane opening angle) and {\it c})~the variable
$E_T(j_1)+E_T(j_2)$.  Signal was found to exceed SM background for
appropriate intervals of each of these variables, for mSUGRA points with
light squarks, where the squark-squark and squark-gluino production
cross-sections are enhanced.

\FIGURE[t]{
\includegraphics[width=9cm,clip]{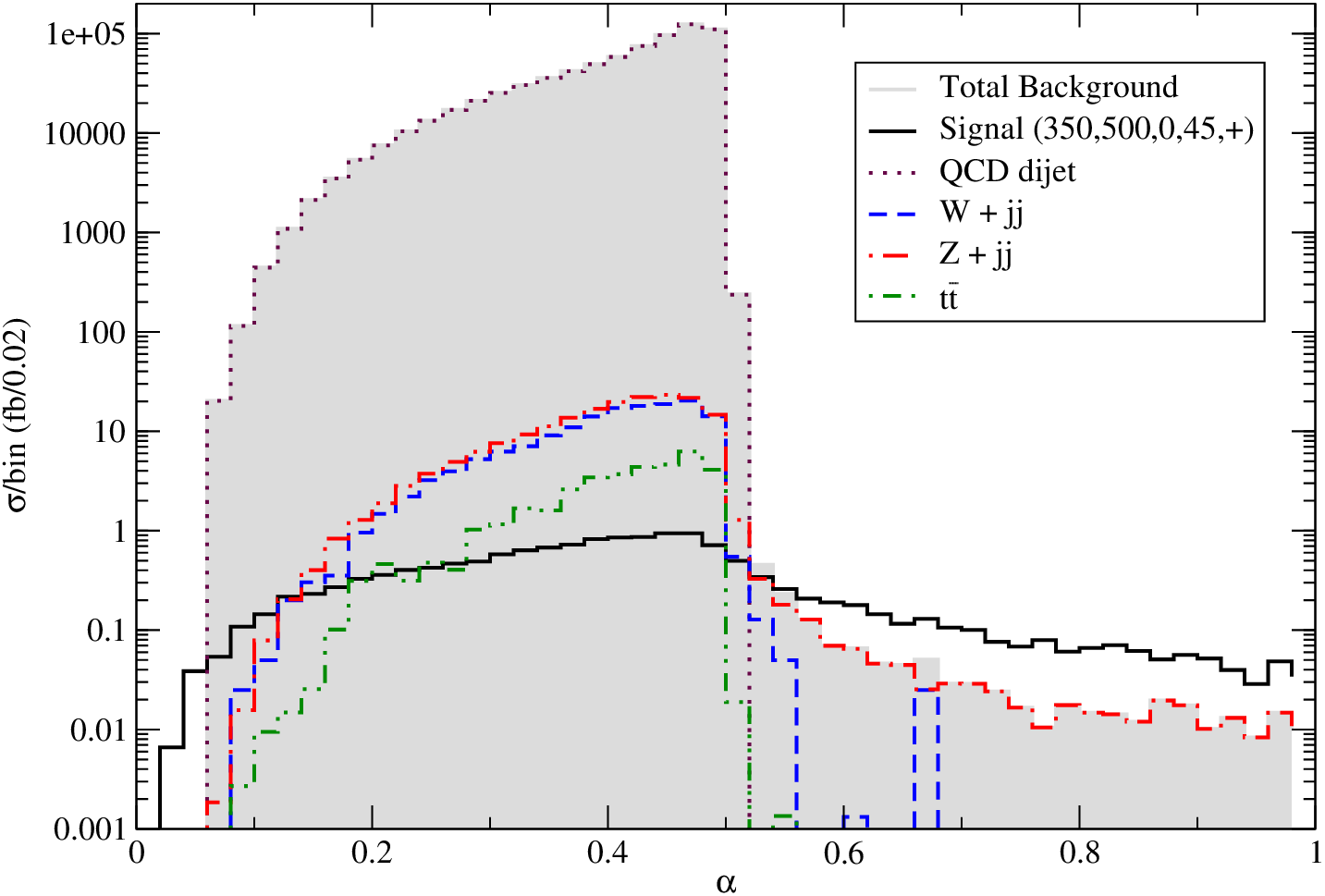}
\caption{ Distribution of $\alpha=E_T(j_2)/m(j_1j_2)$ for dijet 
events with no identified leptons for the mSUGRA
point $m_0=350$ GeV, $m_{1/2}=500$ GeV, $A_0=0$ GeV, $tan\beta=45$ and
$\mu>0$ at $\sqrt{s}=10$ TeV, along with corresponding distributions
from various SM sources. 
The fixed mSUGRA parameters
are $A_0=0$, $\tan\beta =45$ and $\mu >0$.
We require that $E_T(j_1)+E_T(j_2)>700$ GeV,
but make no restriction on $\eslt$.}
\label{fig:dialp}}

Following RT-S, we evaluate signal and background for the LHC start-up
energy of $\sqrt{s}=10$ TeV, and plot the $\alpha$ and $\Delta\phi$
distributions for the mSUGRA point $m_0=350$ GeV, $m_{1/2}=500$ GeV,
$A_0=0$ GeV, $tan\beta=45$ and $\mu>0$.  This mSUGRA point has
moderately heavy gluinos and squarks ($m_{\tilde{g}}=1160$ GeV,
$m_{\tilde{q}}\sim 1000$ GeV) and is {\it not accessible} via early
searches in the multimuon channels.  After imposing a
$E_T(j_1)+E_T(j_2)>700$ GeV cut for both signal and background we see
from Figs.~\ref{fig:dialp} and \ref{fig:diph} that for appropriate cuts
on $\alpha$ and $\Delta\phi$ to remove the enormous QCD background, the
signal is above the remaining background, which mainly comes from the
$Z+2j$ production, where the $Z$ decays invisibly. Dijet events with
the required geometry from $Wjj$ and $t\bar{t}$ production arise
only if one (or more) of the visible decay products of $W$ or $t$ is
missed in the detector. We mention here that for this analysis, we
assume that it will be possible to identify the electron in $W + 2j$
events if $W\to e\nu$, so that such events can be efficiently
vetoed.\footnote{We have also not shown backgrounds from $W+j$ or $Z+j$
production where the $W$ and $Z$ decay to taus that decay
hadronically. Because of the $E_T(j_1)+E_T(j_2)$ cut, these backgrounds
will be significant only for small values of $\alpha$, and peak at large
values of $\Delta\phi$, and will be efficiently removed by the same cuts
that remove the much larger QCD background.}

\FIGURE[t]{
\includegraphics[width=9cm,clip]{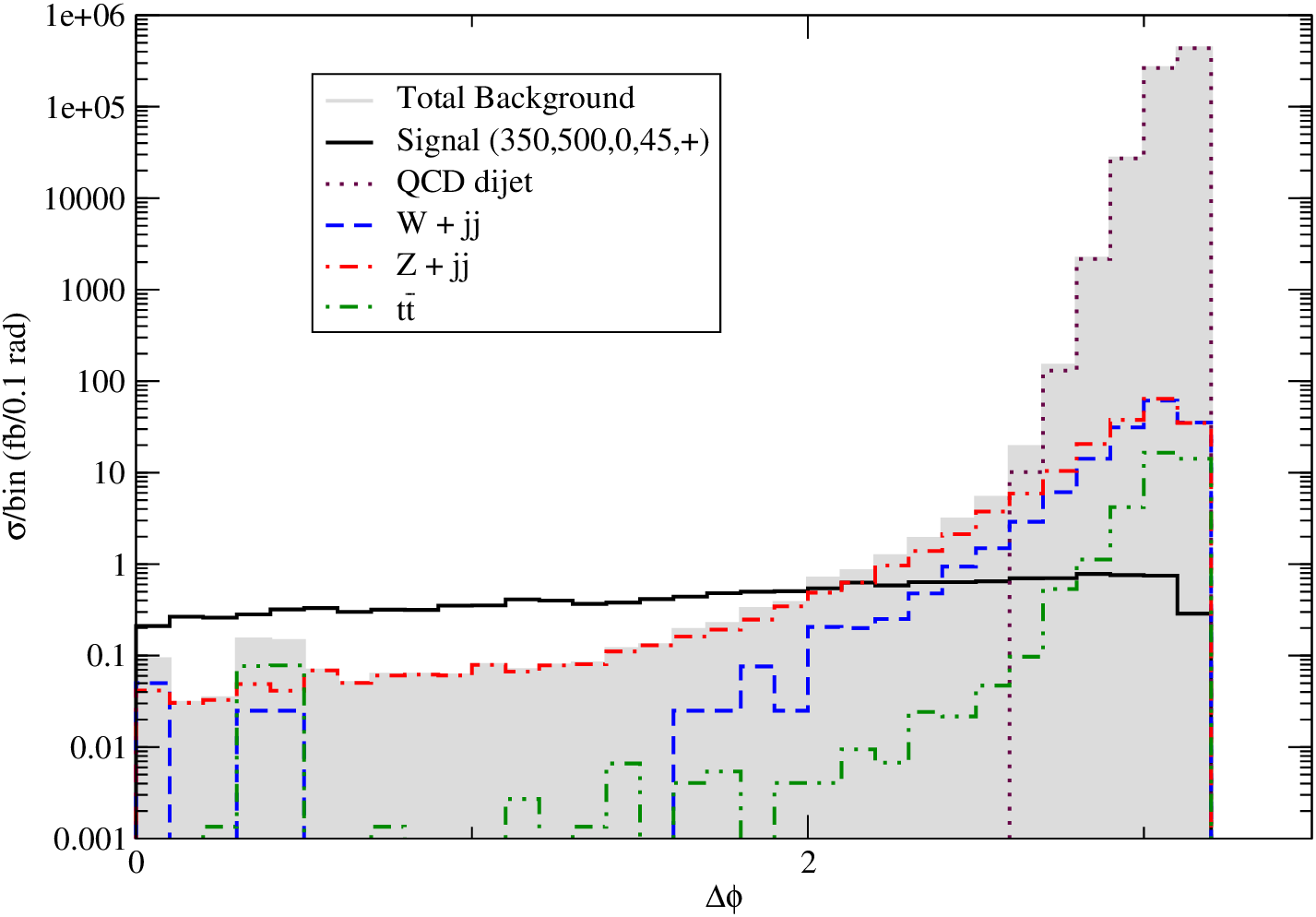}
\caption{ Distribution of $\Delta\phi(j_1,j_2)$ for dijet 
events with no identified leptons for the mSUGRA
point $m_0=350$ GeV, $m_{1/2}=500$ GeV, $A_0=0$ GeV, $tan\beta=45$ and
$\mu>0$ at $\sqrt{s}=10$ TeV, along with corresponding distributions
from various SM sources. We require that $E_T(j_1)+E_T(j_2)>700$ GeV,
but make no restriction on $\eslt$.}
\label{fig:diph}}

To extract the SUSY reach in the dijet
channel, we scan over a large range of mSUGRA model points and find the optimal set of cuts, 
using:
\begin{itemize}
\item $n(jets)= 2$
\item $E_T(j)\ge 50$ GeV
\item $E_T(j_1)+E_T(j_2)\ge 100-1000$ GeV (in steps of 50 GeV)
\item $\alpha> 0.05,0.1-0.9$ (in steps of 0.1)
\item $\Delta\phi(j_1,j_2)< 0.05,0.3-3$ (in steps of 0.3)
\item number of isolated leptons ($\mu$ or $e$), $n(\ell )= 0$
\end{itemize}

The optimal cuts are selected to maximize $S/\sqrt{B+S}$ and satisfy the
discovery criteria defined at the beginning of Sec.~\ref{sec:early}.
Because the important backgrounds all peak at large values of
$\Delta\phi(j_1,j_2)$, the most effective cut is
$\Delta\phi(j_1,j_2)\lesssim2$ over most of the parameter space. Cutting
further on $\alpha$ is then not usually required, so that $\alpha\gtrsim
0.05$ is generally preferred in order to maximize the signal.  As
expected, the optimal $E_T(j_1)+E_T(j_2)$ cut increases with
$m_{1/2}$, but is essentially independent of $m_0$ for
$m_0\lesssim450$ GeV.

Our results for the optimized reach in the dijet channel are shown in
Fig. \ref{fig:rts}. We see that LHC experiments will begin to probe SUSY
in this channel even with just 0.05 fb$^{-1}$ of integrated luminosity,
if systematics are under control. With 1~fb$^{-1}$, LHC
experiments should be able to probe to $m_{1/2}$ and $ m_0$ values
almost up to 500~GeV, corresponding to $m_{\tq}\sim m_{\tg}$ just over 1~TeV!
In the white region at very small $m_0$, the neutralino is not the LSP. As
anticipated, the dijet search mainly probes the small $m_0$ region where
squark pair production forms a significant part of the SUSY cross
section. 
%
\FIGURE[t]{
\includegraphics[width=9cm,clip]{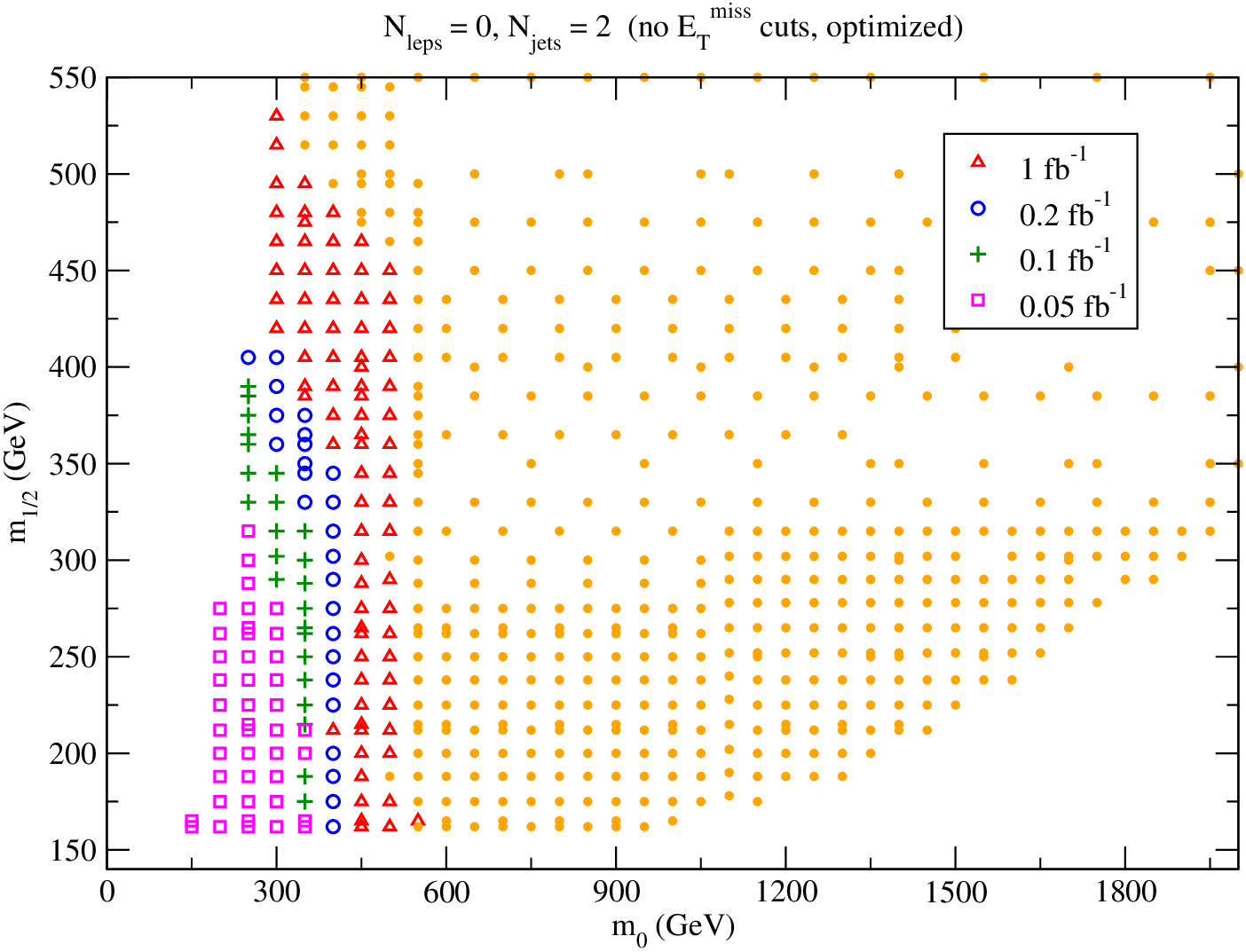}
\caption{ Optimized reach of LHC for mSUGRA at $\sqrt{s}=10$ TeV via the
RT-S dijet search, for various values of integrated luminosity. We assume
that it will be possible to veto events with electrons or muons, but
require no restriction on $\eslt$.
}\label{fig:rts}}
%

\section{Ultimate reach of the LHC utilizing  $\eslt$}
\label{sec:reach}

As the experiments accumulate data, the detectors will become better
understood, and it will be possible to utilize both electrons and
$\eslt$ (well known to be a powerful discriminator between SUSY and SM
events) in the analysis. Because of technical issues to do with
re-training of the magnets, at present there is no clear projection for
the time-line over which the energy of the LHC will be increased from
the initial value of 7-10~TeV to its design value of 14~TeV.  It is clear,
however, that the energy increase will be staged \cite{jenni}. The
possibility that the design energy may not be attained for an extended
period motivated us to study the impact of machine energy as well as
integrated luminosity on the ultimate SUSY reach of the LHC. Toward this
end, we delineate the LHC reach, including electrons and $\eslt$ in the
analysis of the signal, for $\sqrt{s}=10$ and 14~TeV, for various values
of integrated luminosity.  Projections for intermediate
energies may be obtained by interpolating between the reach for these
two extreme values.

Before turning to the results, we draw attention to a potentially
serious problem that arises when we try to make reach projections for
integrated luminosities in the ab$^{-1}$ range. In this case, SM
backgrounds have to be limited to ab levels to make reliable projections
for the observability of a signal (near the five event limit) with
extremely hard cuts on jet and $\eslt$. In spite of the very large Monte
Carlo background samples that we have generated, we are forced to
extrapolate our calculated backgrounds to obtain estimates of these
backgrounds out to high values of $\eslt$ for which our simulation
becomes statistics-limited. For each process and for each set of jet and
lepton cuts listed in  Sec.~\ref{sssec:opt_noetm}, we extrapolate the
background from the lower $\eslt$ range to higher values of $\eslt$
using an exponential fit whenever we have enough events in (at least
three) low $\eslt$ bins to allow a sensible extrapolation. In cases where
this is not possible (mostly $W$ and $Z$ events with multiple jets and
hard jet cuts) we do not extrapolate in this particular channel. However
to avoid a huge under-estimation of this background when sensible
extrapolation is not possible, we first check whether the
non-extrapolated background level (in the channel in question) at low
$\eslt$ exceeds the corresponding extrapolated background cross-section
(often dominantly from $t\bar{t}$ production which, we have checked, can
be reliably extrapolated in most cases) by a factor $\ge 5$.  If it
does, we regard this set of cuts as {\it unsafe} (because the background
that we could not extrapolate may indeed be too large) and exclude them
from our optimization procedure. \footnote{We have checked that dropping
the set of cuts with the large $W/Z$ background does not affect the
reach that we obtain in any significant way. The optimization procedure
picks out a different configuration where the $W/Z$ background is
smaller than the top background.}  We recognize that this procedure may
still underestimate the background at large $\eslt$ in cases where the
non-extrapolated $W$ and $Z$ backgrounds are below five times the total
extrapolated one at low $\eslt$, but may become the dominant source at
large $\eslt$, due to a flatter $\eslt$ spectrum (when compared with the
extrapolated ($t\bar{t}$) one), or simply because we did not obtain any
$W/Z+j$ events in our simulation.   We,
therefore, regard the reach obtained after applying the procedure just
described as the outer limit of the parameter plane that may be probed
for the corresponding integrated luminosity and label it as oFIT
(optimistic fit) in the reach plots below.  
We assume that the QCD background is
smaller than the other backgrounds for high enough values of $\eslt$
\cite{qcd}, so no extrapolation is done in this case. As already noted,
it is frequently possible to extrapolate the $t\bar{t}$ background quite
reliably, and extrapolation of backgrounds from multiple quark and
vector boson production processes (if needed) is even more straightforward.
Finally, we remark that even
for these very large integrated luminosities, we have moderate control
of the extrapolation of the SM backgrounds in the SS dilepton and
$n_{\ell} \ge 3$ lepton channels, where $t\bar{t}$ production (or for
higher lepton multiplicities, multiple vector boson processes) is the
dominant background source and the $W$ and $Z$ backgrounds can be
neglected.  We regard our projection of the reach limited to these
channels as conservative, and label it by cFIT in the plots to follow.

\subsection{LHC reach using $\eslt$ at $\sqrt{s}=10$ TeV} \label{ssec:reach10}
%

We begin by reanalyzing the optimized reach of the LHC at
$\sqrt{s}=10$~TeV. In addition to the optimization over 
$n_j, \ E_T(j_1), \ E_T(j_2)$ as in Sec.~\ref{sssec:opt_noetm},
we now include {\it all} leptons ($e$ and $\mu$) and also optimize 
over, 
%
%
\bi
\item $\eslt\ge 0-1500$ GeV (in steps of 100~GeV).
\item $n(\ell) = 0,1,2,3,4,5,6$ \ei where $\ell=\mu,e$. 
For the $n_{\ell}=2$ OS dilepton signal, we continue to veto events with 
like-flavor, OS dilepton pairs with 75~GeV$\leq m(\ell^+\ell^-)\leq 105$~GeV
or $m(\ell^+\ell^-)\leq 10$~GeV.

\FIGURE[t]{
\includegraphics[width=10cm,clip]{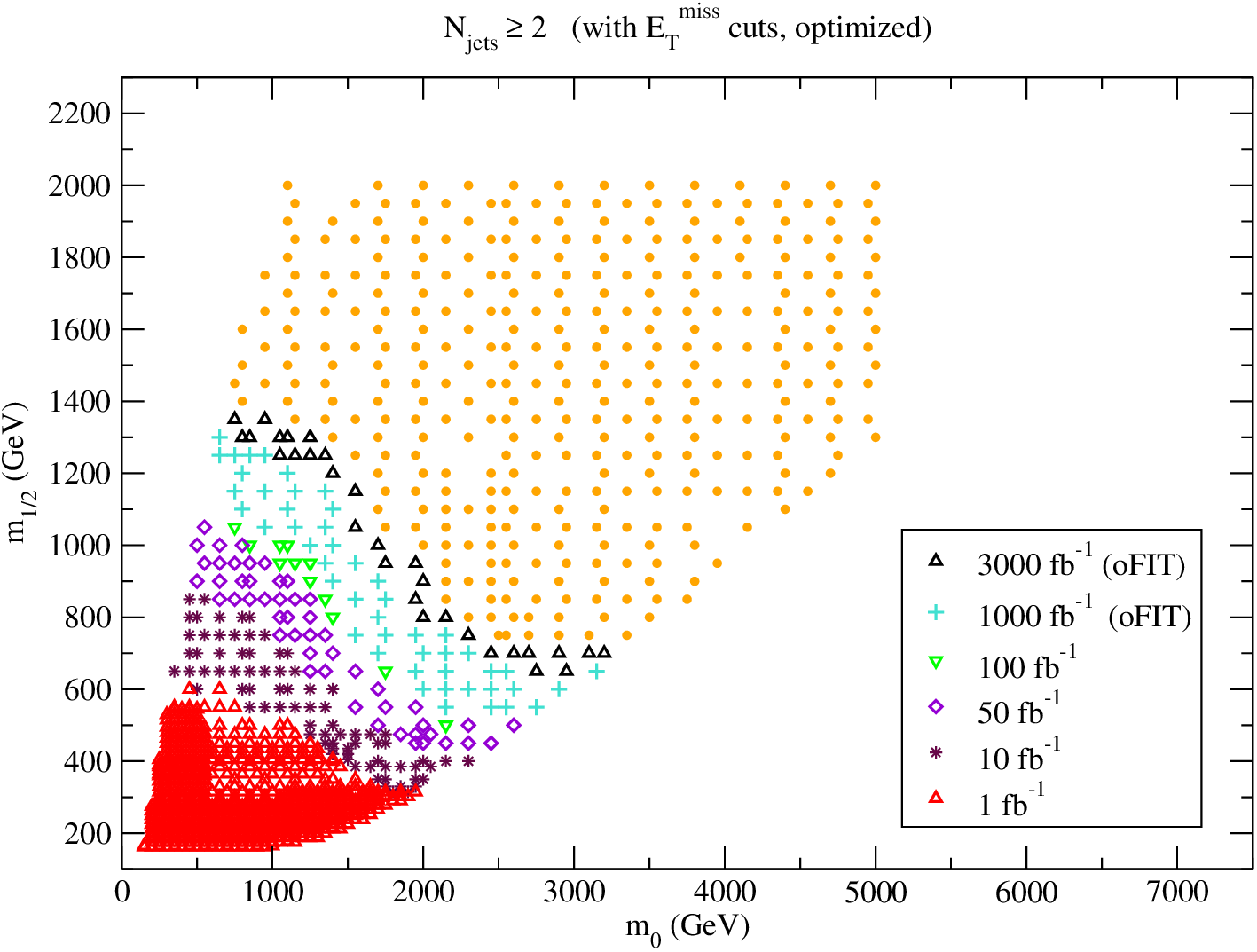}
\includegraphics[width=10cm,clip]{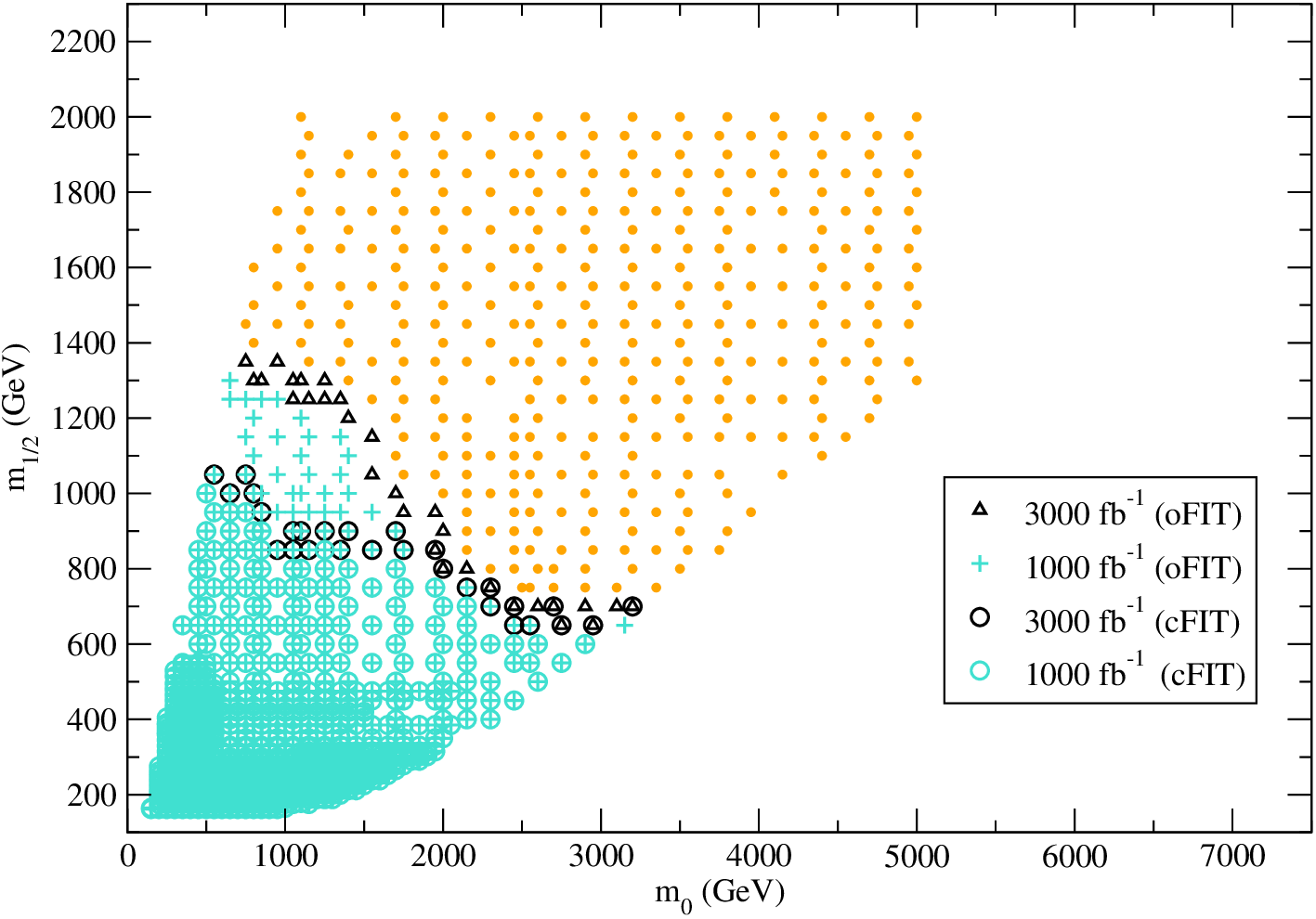}
\caption{ The upper frame shows the reach of LHC for mSUGRA at
$\sqrt{s}=10$ TeV using optimized cuts with $N(jet) \ge 2$, electron ID
and $\eslt$ for various values of integrated luminosities.  For
integrated luminosities of 1000 and 3000~fb$^{-1}$, the reach is
obtained using the oFIT extrapolation procedure (discussed in the text)
that likely underestimates the SM background, and leads to a
correspondingly optimistic projection for the reach. 
The reach in the SS dilepton channel for 1000 and 3000~fb$^{-1}$ 
(labeled cFIT) is
shown by the circles in the lower frame, where the corresponding oFIT
reach is also shown for comparison. We regard this as a conservative
estimate of the SLHC reach.
The fixed mSUGRA parameters
are $A_0=0$, $\tan\beta =45$ and $\mu >0$.  }\label{fig:sug10nge2}}

Our results are shown in the top frame of  Fig. \ref{fig:sug10nge2}.  The
reach in $m_{1/2}$ for low (high) $m_0$ extends to $m_{1/2}=550$
(315)~GeV for an integrated luminosity of 1~fb$^{-1}$, to be compared
with 325 (250)~GeV in Fig.~\ref{fig:sug10}. Thus, $\eslt$, together with
help from electron ID leads to about a 60\% (25\%) increase in the
gluino mass reach if $m_{\tq}\sim m_{\tg}$ ($m_{\tq}\gg
m_{\tg}$).\footnote{We have explicitly checked that the optimization
mostly picks out the $n_{\ell}=0$ channel, allowing us to conclude that
the increased reach is essentially due to the availability of $\eslt$
rather than of the electron signal.}
%
This same analysis gives projections for the machine reach with
increased integrated luminosity. For integrated luminosities of 1, 10
and 100~fb$^{-1}$, the gluino mass reach (for $m_{\tq}\sim m_{\tg}$) is
$\sim$ 1.4, 1.9, 2.3~TeV. For integrated luminosities of 1000 and
3000~fb$^{-1}$, our projection for the {\it outer limit} of the LHC reach
obtained using optimistic extrapolation of the SM backgrounds (oFIT)
described above, extends out to 2.8 and 2.9~TeV, respectively. We
caution that in this region our optimization procedure often selects out the
$n_j=2,\ n_{\ell}=0$ or 1 topology with very hard jet and $\eslt$ cuts, 
where the signal (for the 3000~fb$^{-1}$ case) is just 4~ab. 
At such low cross-sections the signal
is likely termed ``observable'' only because the background from $W/Z+j$
production may be
greatly underestimated. Aside from this, detector issues that will arise
during 
LHC operation at very high instantaneous luminosity have not been
included in this analysis, and may significantly reduce the 
1000~fb$^{-1}$ and 3000~fb$^{-1}$ reach projections. The oFIT
gluino mass reach drops to about 1.3~TeV (1.7~TeV) for an integrated
luminosity of 100 (1000)~fb$^{-1}$ if squarks are very heavy. 
We have checked that, after optimization, the signal almost always arises
from gluino and squark production except at the largest values of $m_0$
where $\tw_2$ and $\tz_4$ production contribute about a third of the signal.
We repeat
that, except possibly at the highest machine luminosities, the ultimate
reach in $m_{1/2}$ in the HB/FP region will be about 15\% higher than
shown in this figure once $b$-jet tagging is utilized to enhance the
signal over the SM background \cite{btag}.\footnote{We mention that we
have not imposed any cut on the transverse mass between the lepton and
$\eslt$ in our analysis of single lepton events. Since this cut is very
efficient at removing backgrounds from $W+j$ and also $t\bar{t}$ events
where just one $W$ decays leptonically, it may be that with an
$m_T(\ell,\eslt) \agt 100$~GeV cut, a slightly bigger reach may be
obtained in the single lepton channel. In the analysis presented here,
the $0\ell$ channel almost always yields the largest reach.}

To obtain a rough idea of how much the oFIT extrapolation may
overestimate the reach, we compare the reach -- shown by circles and
labeled cFIT -- that we obtain in the SS dilepton channel (for which we
have moderate control on the SM backgrounds even for ab$^{-1}$
integrated luminosities) with the oFIT reach in the lower frame of
Fig. \ref{fig:sug10nge2}. We see that for small values of $m_0$, the
envelope of even the black circles, roughly speaking, follows the
100~fb$^{-1}$ reach triangles in the upper frame, but for large $m_0$
gives a somewhat increased reach in $m_{1/2}$. Once again, except at the
highest values of $m_0$ where $\tw_2$ and $\tz_4$ production contributes
up to a quarter, the signal (after optimization) is dominated by squark
and gluino production. Modulo detector issues at high luminosity, we
regard the cFIT reach shown in the figure as a conservative projection
of the SLHC reach.


\subsection{LHC reach with $\eslt$ at $\sqrt{s}=14$ TeV}

We repeat the optimized cuts analysis that led to the reach in
Fig.~\ref{fig:sug10nge2}, for $\sqrt{s}=14$ TeV.  As for the 10~TeV case
just discussed, we first show results using the oFIT to the backgrounds
for SLHC integrated luminosities in the upper frame of
Fig. \ref{fig:sug14}.  We see that the corresponding gluino mass reach
(for $m_{\tq}\sim m_{\tg}$), shown in Table~\ref{table:reach}, extends
to $m_{\tg}$ = (2.4, 3.1, 3.7)~TeV for an integrated luminosity of (10,
100, 1000)~fb$^{-1}$. We see that the typical gain in the gluino mass
reach due to the increased machine energy is a factor 1.3-1.4, {\it
i.e.} this reach, roughly speaking, scales with the center-of-mass
energy, and extends to over 3~TeV (4~TeV) for an integrated luminosity
of 100 (3000)~fb$^{-1}$.  We have checked that at the highest $m_{1/2}$
and modest $m_0$ values for which we find an observable signal our
optimization procedure again picks out the $n_j=2$ channel with very
hard jet and $\eslt$ cuts with a signal cross section of $\sim 2$~ab,
and for the reasons mentioned in Sec.~\ref{ssec:reach10}, the oFIT reach
shown is almost certainly an overestimate.

\FIGURE[t]{
\includegraphics[width=9cm,clip]{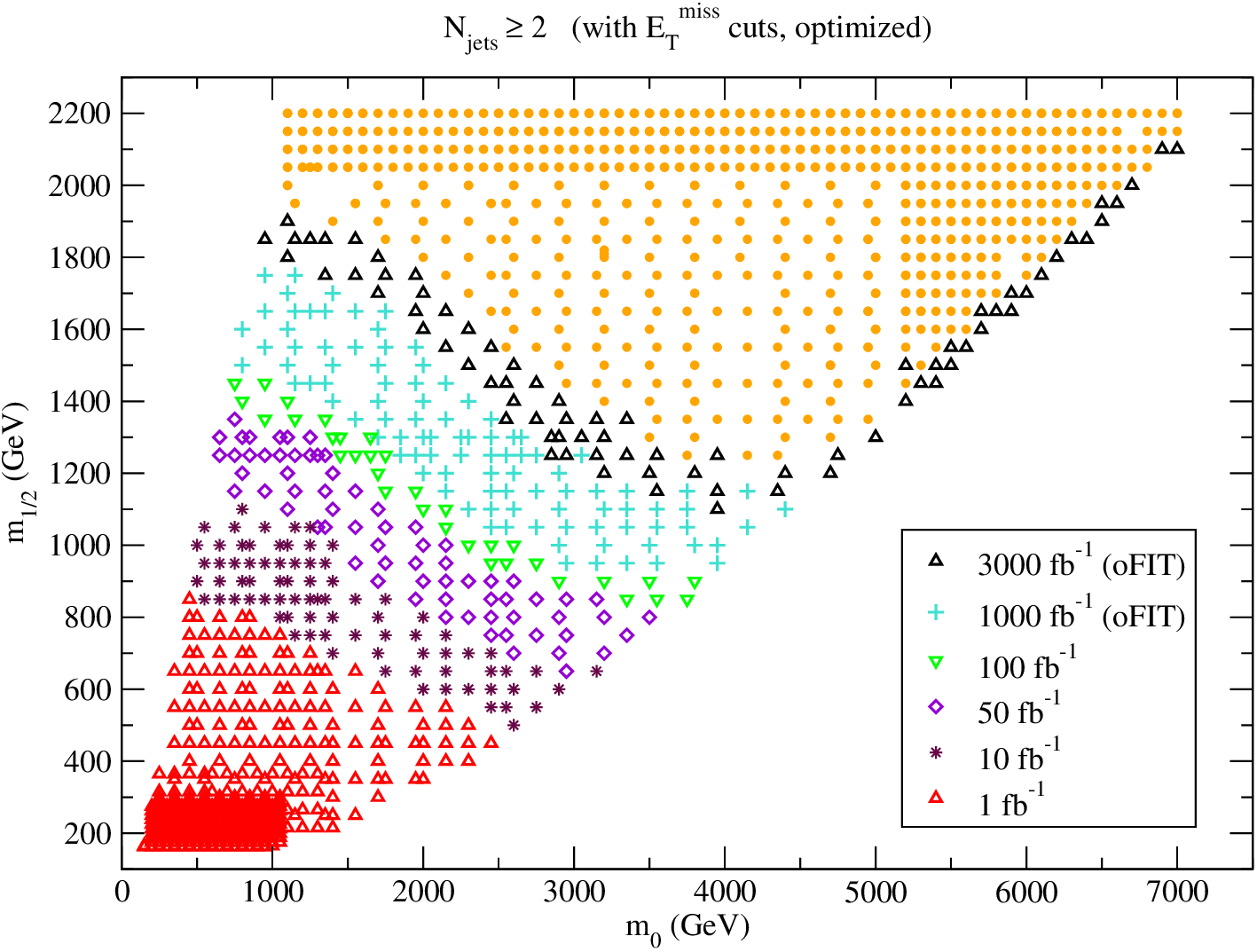}
\includegraphics[width=9cm,clip]{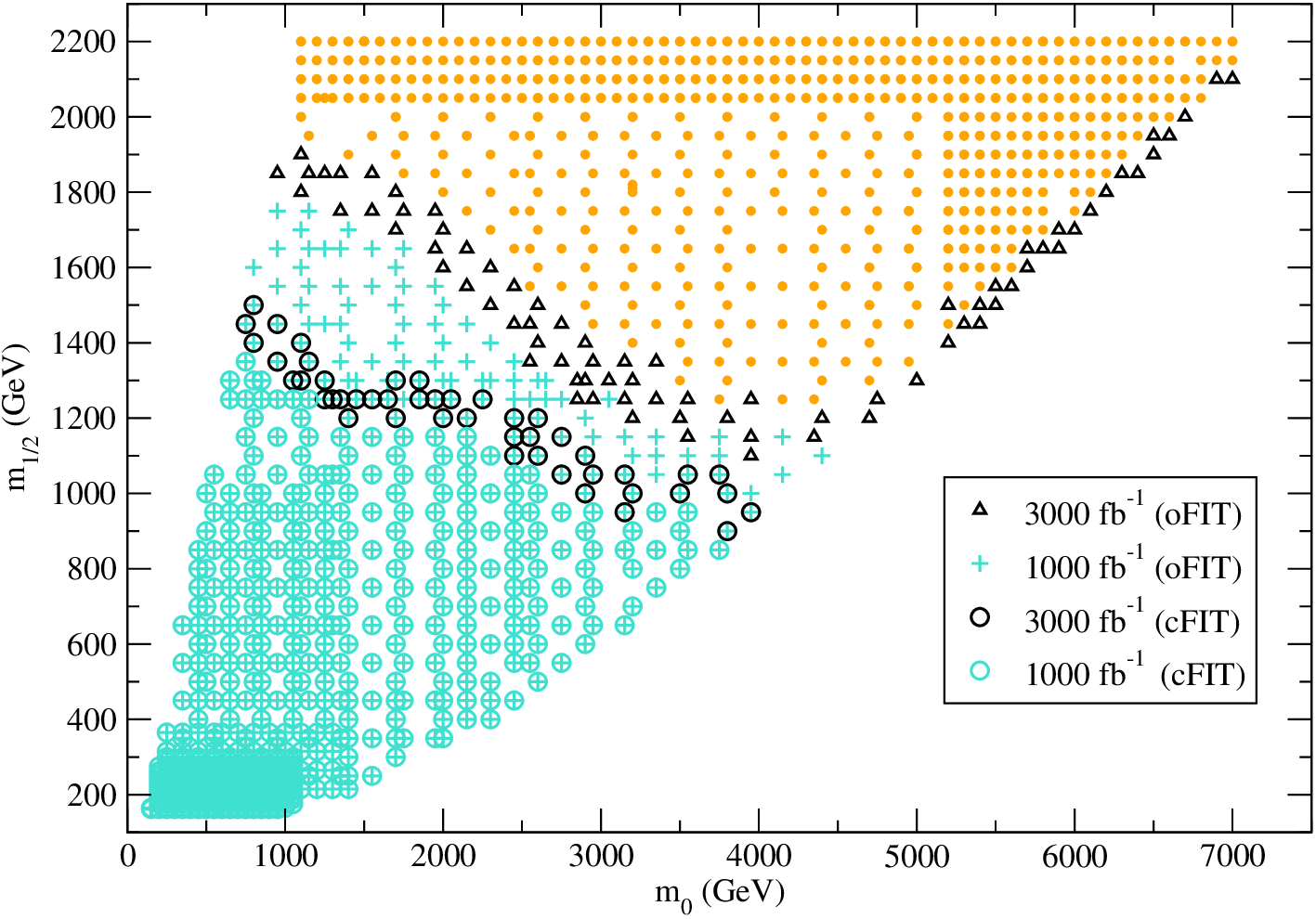}
%
\caption{ The upper frame shows the reach of LHC for mSUGRA at
$\sqrt{s}=14$ TeV using optimized cuts with $N(jet) \ge 2$, electron ID
and $\eslt$ for various values of integrated luminosities.  For
integrated luminosities of 1000 and 3000~fb$^{-1}$, the reach is
obtained using the oFIT extrapolation procedure (discussed in the text)
that likely underestimates the SM background, and leads to a
correspondingly optimistic projection for the reach. 
The reach in the SS dilepton channel for 1000 and 3000~fb$^{-1}$ (and
labeled cFIT) is
shown by the circles in the lower frame, where the corresponding oFIT
reach is also shown for comparison. We regard this as a conservative
estimate of the SLHC reach.
The fixed mSUGRA parameters
are $A_0=0$, $\tan\beta =45$ and $\mu >0$.  }\label{fig:sug14}}

The most striking new feature in the upper frame is the appearance of a
signal in the HB/FP region for an integrated luminosity of
3000~fb$^{-1}$. We have checked that the signal in this region arises
only from $\tw_2$ and $\tz_4$ production, in the $n_j\ge 2$ channel with
very hard jet and $\eslt$ cuts. Even in the region near $m_0\sim
4000$~GeV and $m_{1/2}=1200$~GeV where the SLHC reach contours flatten
out, electroweak-ino production accounts for about half the signal,
whereas for yet lower values of $m_0$, the signal originates essentially
in gluino and squark events. The signal deep in the HB/FP region is just above
the $5\sigma$ level with a cross-section of $\sim$5 ab and,
for fixed $m_{1/2}$, rapidly falls if $m_0$ is reduced 
because the mass gap between the wino
parent and the higgsino LSP quickly reduces, softening the $\eslt$
spectrum, concomitantly reducing the efficiency for the signal to pass
the hard $\eslt$ cut required to reduce the background. Indeed, it is
very likely that we obtain the reach in the HB/FP region only because
we get no background from $W/Z+j$ production  in one (or more)
of the many channels  examined in the course of the optimization.

In the lower frame of Fig.~\ref{fig:sug14} we show the corresponding
reach in the SS dilepton channel. As in the previous figure, the cFIT
circles denote points where the signal is observable in the SS dilepton
channel. Except in the region at $m_{1/2}\sim 1$~TeV and $m_0> 3$~TeV
where about half the signal comes from wino production, the SS dilepton
signal arises mostly from gluino and squark production. We also see that
the HB/FP region entirely disappears. Since any signal in this region is
most likely to come from $\tw_2$ and $\tz_4$ production which yields SS
dilepton events only if at least one final state lepton fails to be
identified in the detector, we also checked that there is no observable
signal in the $n_{\ell} \ge 3$ lepton channel, or for that matter in
the inclusive SS and $n_{\ell}\ge 3$ event channels. We see that as in the
$\sqrt{s}=10$~TeV case, our more conservative projection for the reach is
again close to the projected reach with 100~fb$^{-1}$ shown in the upper
frame of the figure.

\section{Concluding remarks and summary }
\label{sec:conclude}

LHC experiments will soon probe particle collisions in a qualitatively
new energy regime and, we hope, uncover new phenomena. In this paper, we
access the LHC reach for supersymmetric particles, both at the very
early stages of LHC operation at the starting center-of-mass energy of
10~TeV and limited values of integrated luminosity (0.1-1~fb$^{-1}$)
when the detectors will not yet be completely understood and calibrated,
as well as at later stages of LHC operation at $\sqrt{s}=10$ and 14~TeV,
again for several values of integrated luminosity extending up to the design
value, and beyond into the super-LHC stage.  To obtain these reach
projections, we have used improved techniques described in
Sec.~\ref{sec:bg} to calculate SM backgrounds from a large number of
$2\to n$ processes.

It has, however, often been stated that exploration of new physics
(except for resonances that can be reconstructed as mass bumps) will
only be possible after all the detectors systems are completely
understood, and the ``SM is rediscovered at the LHC''. This dictum has
been thought to be especially true for the discovery of
supersymmetry\footnote{The dictum applies equally well to a variety of
proposals that address the mechanism of electroweak symmetry breaking
and the stabilization of the electroweak scale and
include a stable weakly interacting massive particle that escape
detection in the experimental apparatus \cite{hb_tasi}.}  since the experimental
determination of $\eslt$, which is an essential element of the canonical
SUSY signal, truly does require a reliable identification and
measurement of all high $E_T$ electrons, muons, photons and jets in each
event. This motivated us to examine the prospects for early sparticle
detection without use of $\eslt$.

Since  electron fakes from jets could be a serious issue in the early
stages of operation, and since non-leptonic and single-muon signals will
have large SM backgrounds, in Sec.~\ref{sec:early} we focused on multi-muon
 signals from SUSY without any requirement on $\eslt$. We concluded
that at the LHC start-up energy $\sqrt{s}=10$~TeV, the OS dimuon channel
(which has the highest rate of the multi-muon signals) offers the best
prospects for the earliest detection of SUSY: even with very basic cuts and
just 100~pb$^{-1}$ of integrated luminosity, LHC experiments will probe
significant portions of mSUGRA parameter space beyond the range of
Tevatron experiments. The dimuons from the SUSY signal will have masses
that tend to cluster close to, but below the expected mass edge at
$m_{\tz_2}-m_{\tz_1}$.  As more data are accumulated, the relatively
background-free but rate-limited SS dimuon and trimuon channels
become the more important, with the latter yielding the highest reach,
with gluinos as heavy as 700~GeV being accessible with 1~fb$^{-1}$ of
data if $m_{\tq}\sim m_{\tg}$. This reach may be extended to 800~GeV
using optimized cuts, though by the time such analyses can be performed,
it is possible that
reliable measurement of $\eslt$ will also be available. We have also
presented a number of muon distributions that serve to characterize the
signal, and allow us to make at least a circumstantial case for its
supersymmetric origin. In this connection, see especially the discussion
of the low mass dimuon in trimuon events, and the distribution of the
transverse plane opening angle for OS and SS dimuon events.

\FIGURE[t]{
\includegraphics[width=9cm,clip]{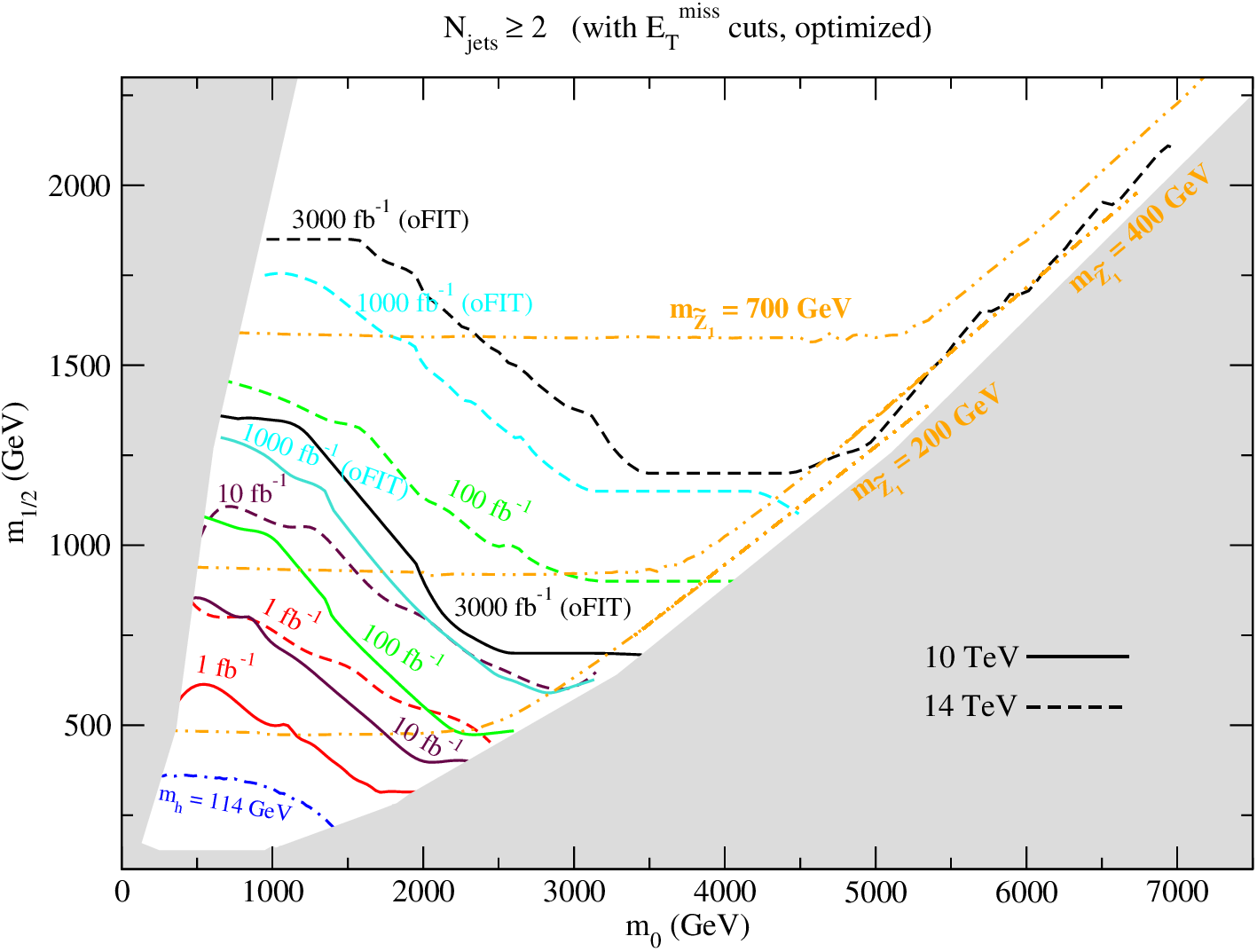}
\caption{ The ultimate SUSY reach of LHC within the mSUGRA framework for
$\sqrt{s}=10$ TeV (solid) and $\sqrt{s}=14$ TeV (dashed) for various
values of integrated luminosities. For integrated luminosities of 1000
and 3000~fb$^{-1}$, we have shown our (over)-optimistic projections
obtained using the oFIT procedure that tends to underestimate SM
backgrounds. A conservative projection for the corresponding reach
essentially follows the 100~fb$^{-1}$ contour.  The fixed mSUGRA
parameters are $A_0=0$, $\tan\beta =45$ and $\mu >0$.  Isomass contours
for the LSP (double dot-dashed) and for a $114$ GeV light Higgs scalar
(dot-dashed) are also shown. The shaded areas are excluded either
because the neutralino is not the LSP, or electroweak symmetry breaking
is not correctly obtained. }\label{fig:curves}}

If squarks are light ($m_{\tq} \sim m_{\tg}$), LHC experiments will be
able to probe squark and gluino production in the acollinear dijet
channel (mostly from squark pair production with squarks decaying
directly to the LSP), {\it without the use of $\eslt$} \cite{rts} out to
masses of several hundred GeV with just 200~pb$^{-1}$ of integrated
luminosity, and to $\agt 1$~TeV with 1~fb$^{-1}$.  We conclude that
within the mSUGRA framework, {\it if squarks and gluinos are in the few
hundred GeV range, there could be a variety of multi-muon and jet
signals that will be detectable in LHC experiments even during the
first run} that is expected to accumulate up to $\sim$200~pb$^{-1}$ of
integrated luminosity, despite the fact that the detectors may not be
well enough understood to allow the use of $\eslt$ or electron signals
in these analyses.

In Sec.~\ref{sec:reach}, we examined the ultimate reach of the LHC for
various integrated luminosities, assuming that the detectors are fully
understood. We show results for $\sqrt{s}=10$ and 14~TeV in
Fig.~\ref{fig:sug10nge2} and Fig.~\ref{fig:sug14}, respectively.  These
are succinctly summarized in Fig.~\ref{fig:curves} where, for 
integrated luminosities of 1000 and 3000~fb$^{-1}$, we have shown
likely over-optimistic reach projections 
obtained using the oFIT extrapolation to
estimate SM backgrounds with very hard cuts. We observe the
following luminosity and energy scaling rules for the approximate
sparticle reach (using optimistic oFIT projections) when $m_{\tq}\sim
m_{\tg}$.

\bi
\item The reach scales with machine energy, so that the gluino reach
  increases by about 40\% between $\sqrt{s}=10$~TeV  and
  $\sqrt{s}=14$~TeV. 

\item At $\sqrt{s}=10 \ (14)$~TeV, every order of magnitude increase in
  luminosity gives an increase in reach of $\sim 400$ ($\sim
  600$)~GeV. We caution though that the oFIT projections very likely
  over-estimate the reach for integrated luminosities at the  ab$^{-1}$
  level, and the growth of the reach with luminosity may well slow down
  after $\sim$100~fb$^{-1}$.
  
\ei

The situation is more complicated for very large values of $m_0$ where
signals from electroweak $\tw_2\tw_2$ and $\tw_2\tz_4$ production appear
to yield an increased reach in the HB/FP region at $\sqrt{s}=14$~TeV and
ab$^{-1}$ values of integrated luminosity as seen by the up-turn of the black
dashed reach contour in Fig.~\ref{fig:curves}. As mentioned earlier, it
is very likely that the oFIT procedure seriously under-estimates the
background from $W/Z+j$ production, and that there is really no reach in
the HB/FP region.  More generally, barring clever new strategies (for
instance, using gauge boson polarizations) to enhance the new physics
signal, we believe that the oFIT contours yield over-optimistic
projections over most of the $m_0-m_{1/2}$ plane.

%
%

Fig.~\ref{fig:gln1} presents a snapshot of the ultimate gluino mass
reach of the LHC, with $\sqrt{s}=10$ (solid histogram) and 14~TeV
(dashed histogram), for several values of integrated luminosity.  The
hatched portions of the bar show the LHC reach in terms of
$m_{\tg}$. The height of the lower hatched portion (bottom-right to
top-left hatching) of each bar shows the value of $m_{\tg}$ up to which
discovery of gluinos is guaranteed at the LHC irrespective of the squark
\FIGURE[t]{
\includegraphics[width=9cm,clip]{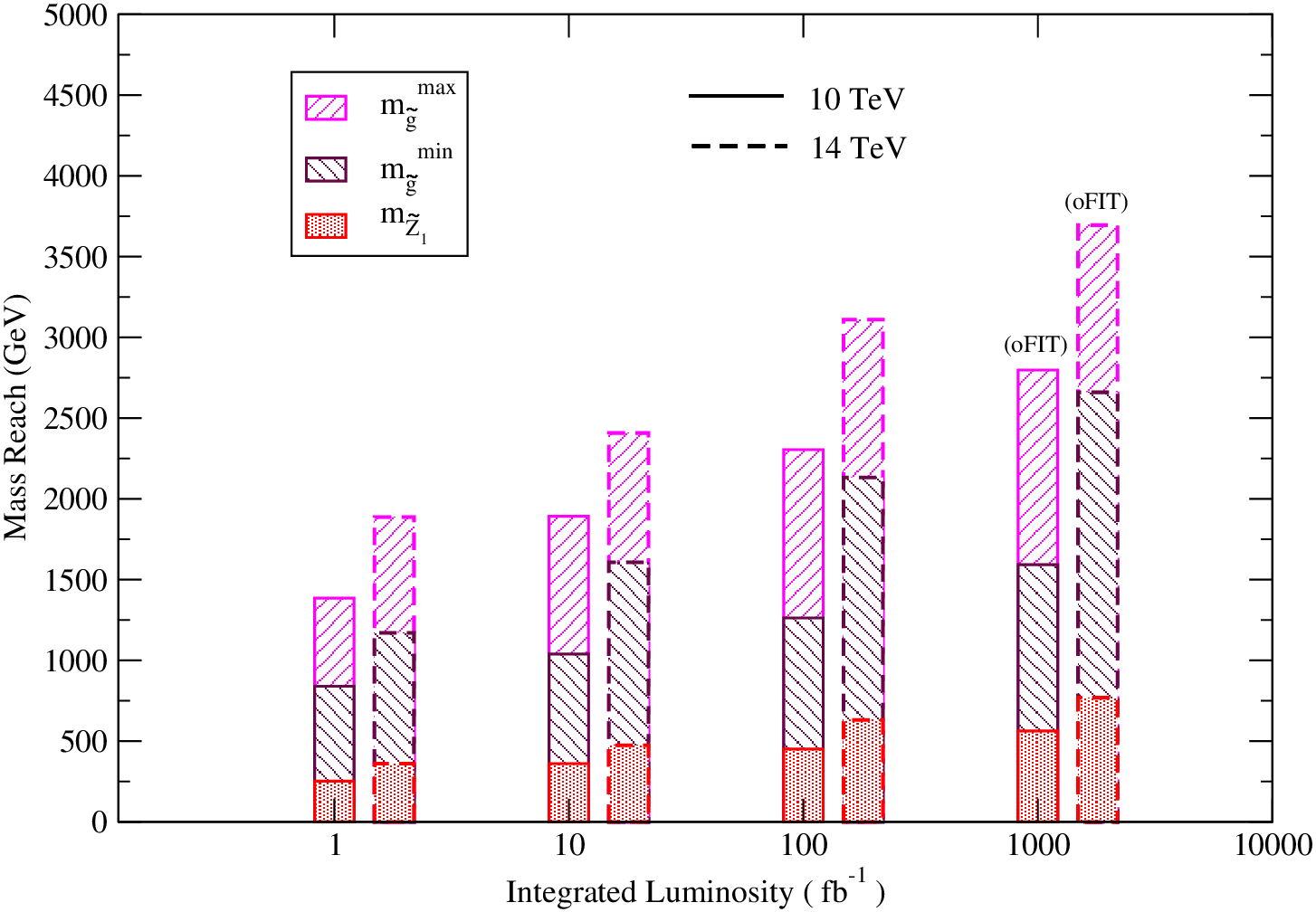}
\caption{The ultimate reach of the LHC at
$\sqrt{s}=10$ TeV (solid) and $\sqrt{s}=14$ TeV (dashed) in terms
of the gluino
and the LSP masses within the mSUGRA framework, for several values of
the integrated luminosity. Results for the 1000~fb$^{-1}$ case are
obtained using the oFIT procedure to extrapolate the background. A
conservative analysis would give results close to those for
100~fb$^{-1}$. 
The heights of the lower (top-left to bottom-right) hatched bars show
the value of $m_{\tg}$ up to which gluinos are guaranteed to be
detectable at the LHC regardless of squark masses, while the heights of
the upper (top-right to bottom-left) hatched bars show the greatest
gluino mass that may be accessible, usually when $m_{\tq} \sim m_{\tg}$.  The
dotted bars show the range of the lightest neutralino mass over the part
of mSUGRA parameter space for which there is an observable SUSY signal
at the LHC.  }\label{fig:gln1}}
mass, while the height of the upper hatched part (top-right to
bottom-left hatching) bars correspond to the maximum value of $m_{\tg}$
that LHC experiments will be able to probe for some value of $m_0$
(usually small), where typically $m_{\tq}\sim m_{\tg}$.  For integrated
luminosities $\ge 1$~ab$^{-1}$, the reach shown is obtained from the likely
over-optimistic oFIT procedure to extrapolate the background.
Finally, the range of the lightest neutralino mass over the region of
mSUGRA parameter space where LHC experiments will be able to detect a
SUSY signal is shown by the red-dotted bars in the figure. Comparing
with earlier studies \cite{dm_det}, we see that there will be detectable
signals in the next round of direct dark matter searches -- XENON, LUX
with $\sim 100$~kg of noble liquids, or superCDMS (25kg) -- over the
entire $\tz_1$ mass range in the figure, and also at IceCube if
$m_{\tz_1} \alt 550-600$~GeV, if parameters are in the HB/FP region,
where the neutralino composition is adjusted to give the measured amount
of dark matter. Instead, for a bino-like $\tz_1$, direct searches with
ton-sized detectors will be sensitive to $\tz_1$ masses up to about
300~GeV.  It is exciting that if supersymmetry is realized as in the
mSUGRA framework (or one of its variants with non-universal parameters),
we expect observable signals not only at the LHC, but also in a
completely different program of experiments unrelated to accelerator
particle physics.

\acknowledgments
We thank M. Drees for many very helpful comments on an early version of
the manuscript.
VB thanks the Aspen Center for Physics for hospitality. 
This research was supported in part by the U.S. Department of Energy, 
by the Fulbright Program and CAPES (Brazilian Federal Agency for Post-Graduate Education).

%

\end{document}